\documentclass[epj]{svjour}
\usepackage{graphics}
\usepackage{hyperref}
\usepackage{amsmath,amssymb}
\usepackage{multirow}
\usepackage{tikz} %test https://texample.net/tikz/examples/commutative-diagram-tikz/
\usepackage{stackrel}
\usetikzlibrary{matrix} %test
\hypersetup{
    colorlinks=true,
    linkcolor=blue,
    filecolor=blue,      
    urlcolor=blue,
    citecolor=blue}
\begin{document}
\title{On the canonical equivalence between Jordan and Einstein frames}
\author{Gabriele Gionti, S.J.\inst{1,2,3} \and Matteo Galaverni \inst{1,4}% etc
}                     % Do not remove
\institute{
Specola Vaticana (Vatican Observatory), V-00120 Vatican City, Vatican City State \and
Vatican Observatory Research Group, Steward Observatory, The University Of Arizona,
933 North Cherry Avenue, Tucson, Arizona 85721, USA  \and 
INFN, Laboratori Nazionali di Frascati, Via E. Fermi 40, I-00044 Frascati, Italy \and 
INAF/OAS Bologna, via Gobetti 101, I-40129 Bologna, Italy}
\mail{ggionti@specola.va}
\date{Received: date / Revised version: date}
\abstract{
A longstanding issue is the classical equivalence between the Jordan and the Einstein frames,
which is considered just a field redefinition of the metric tensor and the scalar field.  
In this work, based on the previous result that the Hamiltonian transformations from the Jordan to the Einstein frame are not canonical on the extended phase space, we study the possibility of the existence of canonical transformations. We show that on the reduced phase space -- defined by suitable gauge fixing of the lapse and shifts functions -- these transformations are Hamiltonian canonical. Poisson brackets are replaced by Dirac's brackets following the Bergman-Dirac's procedure. The Hamiltonian canonical transformations map solutions of the equations of motion in the Jordan frame into solutions of the equations of motion in the Einstein frame. 
\PACS{
      {PACS-key}{discribing text of that key}   \and
      {PACS-key}{discribing text of that key}
     } % end of PACS codes
} %end of abstract
\maketitle
\section{\label{intro} Introduction}
Dicke, in a pioneering article \cite{Dicke}, stressed that physics is invariant under re-definition of unit of measurement. This very fact implies that physics should be invariant under Weyl (conformal) transformation of the metric coefficients. 
The starting frame, where we consider the metric tensor, is called {\it Jordan frame} (JF), while the frame obtained by the Weyl (conformal) transformation of the original metric is called {\it Einstein frame} (EF) \cite{Faraoni2006,FaraoniNardone,Dyer}.
Many people believe that the passage from the Jordan to the Einstein frames is only a field redefinition at classical level
\cite{Cho1992,Deruelle2010,Francfort2019,Frion2018,fluid1,fluid2,fluid3}
as well as at quantum level \cite{Ohta2017,ChristianS2017,Ruf2017,Karamitsos:2018lur,Finn:2019aip,Mukherjee:2023qan}.  
In this last case, there are people claiming that the two frames are inequivalent  \cite{Falls2018,Kamenshchik2014,Filippo2013,Pandlejee2016}.
The equations of motion in the Jordan frame have been found completely equivalent to those one in the Einstein frame \cite{Capozziello:1996xg,Tsamparlis:2013aza,Capozziello2010,Carloni:2007eu,Carloni:2010rfq}.

In order to show a concrete example, we consider a special case of scalar-tensor
Brans--Dicke theory \cite{Brans1961}.
In the {\it Jordan frame} the action is \cite{Dyer}: 
\begin{eqnarray}
S&=&\int_{M}d^{4}x\sqrt{-g}\left(\phi\;{}^{4}R-\frac{\omega}{\phi}g^{\mu\nu}\partial_{\mu}\phi \partial_{\nu} \phi -U(\phi)\right)\nonumber \\
&&+ 2\int_{\partial M} d^3x \sqrt{h}\phi K\;.
\label{BDaction}
\end{eqnarray}
The equations of motion for the metric tensor are:
\begin{eqnarray}
R_{\mu \nu }-\frac{1}{2}g_{\mu \nu }R&=&\frac{\omega}{\phi^2}\left[\partial_\mu\phi\partial_\nu\phi-\frac{1}{2}g_{\mu\nu}g^{\alpha\beta}\partial_\alpha
 \phi \partial_\beta \phi)\right]+\nonumber\\
 &+&\frac{1}{\phi}\left[\nabla_\mu\nabla_\nu\phi-g_{\mu\nu}\Box \phi -\frac{1}{2}g_{\mu\nu}U(\phi)\right] ,\nonumber\\
 \label{equationforg}
\end{eqnarray}
and for the scalar field:
\begin{equation}
(3+2\omega)\Box \phi =\phi \frac{dU}{d\phi}-2U(\phi)\,.
\label{eqstophi}
\end{equation}

Weyl (conformal) transformations of the metric:
\begin{eqnarray}
{\widetilde g}_{\mu\nu}&=&(16\pi G \phi)\, g_{\mu\nu}\,,\;\;{\widetilde \phi}=\phi\,,
\label{Weyltrans1}
\end{eqnarray}
define the passage from Jordan frame, see Eq.~\eqref{BDaction}, to {\it Einstein frame}: 
\begin{eqnarray}
{\widetilde S}&=&\int_{M}d^{4}x{\sqrt{-{\widetilde g}}}\left[ \frac{1}{16\pi G}{\widetilde R}
-\frac{2\omega+3}{(16\pi G)2\phi^2}{\widetilde g}^{\mu\nu}\partial_{\mu}\phi\partial_{\nu}\phi\right. \nonumber\\ &&
\left. -V(\phi)\right]
+\frac{1}{8\pi G}\int_{\partial M}d^{3}{\sqrt{\widetilde h}}{\widetilde K}
\label{scalartensorEF},
\end{eqnarray}
where: 
\begin{equation}
V(\phi)\equiv\frac{U(\phi)}{(16\pi G\phi)^2}\,.
\label{Vtilde}
\end{equation}

Equivalence of the Jordan and the Einstein frames means that if the following couple 
\begin{equation}
\left (g_{\mu\nu}(x), \phi(x) \right)
\label{primacoppia}
\end{equation}
is solution of the equations of motion \eqref{equationforg}-\eqref{eqstophi} in the Jordan frame, then the corresponding couple, obtained through Weyl (conformal) transformation 
\begin{equation}
\left ({\widetilde g}_{\mu\nu}(x), \phi(x) \right)
\label{primacoppia2}
\end{equation}
is solution of the equations of motion derived from the action \eqref{scalartensorEF} in the Einstein frame. One way for proving the previous statement is to go to the Hamiltonian formalism and show that the transformations from the Jordan to the Einstein frames are Hamiltonian canonical. 

In the literature several authors 
claimed \cite{Garay:1992ej,Ezawa1998,Barreto2017}, or partially proved \cite{Deruelle2009,Kluson:2022qxl}, that the Hamiltonian transformations from the Jordan to the Einstein frame are canonical transformations. 
Therefore the Dirac's constraint analysis either of the Brans-Dicke theory \cite{Garay:1992ej} or a generic $f(R)$ theory has been carried out making a transformation from the Jordan to the Einstein frame. In virtue of the assumed canonicity of the Hamiltonian transformations from the Jordan to the Einstein frames, the constraint algebra of the secondary first class constraints in the Jordan frame is just the same of Einstein's geometrodynamics \cite{Kuchar1}. 
We showed, see \cite{Gionti2021,Galaverni:2021xhd,Galaverni:2021jcy}, that the Hamiltonian transformations from the Jordan to the Einstein frames cannot be considered canonical transformations strictly speaking.  
In the Einstein frame there are Poisson brackets, among ``non-conjugate'' variables, which are not zero \cite{Gionti2021,Galaverni:2021jcy,Galaverni:2021xhd}: 
\begin{equation}
\{{\widetilde N}(x),{\widetilde \pi}_{\phi}(x') \}=\frac{8\pi GN(x)\delta^{(3)}(x-x')}{{\sqrt{16 \pi G\phi(x)}}} \neq 0\,.
\label{non-commuto}
\end{equation}

As it is well known, the lapse $N$ and the shifts $N_{i}$ are gauge variables related to the coordinate displacements between two space-like surfaces. We have shown that also in the Hamiltonian formalism of mini-superspace model of flat FLRW universe the transformations from the Jordan to the Einstein frames are not canonical transformations \cite{Galaverni:2021xhd}. 
As a further check of non-canonicity, we have transformed the Hamiltonian function from the Jordan to the Einstein frames. 
We derived the equations of motions both in the Jordan and in the Einstein frames using the respective Hamiltonians. 
In the Einstein frame, on the equations of motions, we applied the transformations from the Einstein to the Jordan frames. In this way, we get two sets of equations of motions in the Jordan frame. We confronted and contrasted these two sets and found that all the equations of motion are equivalent {\it modulo} constraints, except the equation of motion for the lapse function. 
This very fact suggested us to gauge-fix the lapse function both in the Jordan frame and in the Einstein frame. 
We implemented the gauge-fixing condition as secondary constraint and noticed, as expected, that these secondary constraints become second class constraints with the primary constraints. Introducing Dirac's brackets, we derived the equation of motions and solved strongly the second class constraints. 
Therefore, we end up with a reduced phase space in which the lapse N and the related momentum are not a dynamical variables anymore. 
The transformations from the Jordan to the Einstein frames, restricted on this reduced phase-space, is now Hamiltonian canonical. 
The same reasoning can be implemented in the field theory case. There, we have to gauge-fix the lapse $N$ and the shifts $N^{i}$.
The transformations from the Jordan to the Einstein frame continue to be Hamiltonian canonical.

The paper is organized as follows: in Sec.~\ref{Sect:finite}
we study the Hamiltonian Weyl (conformal) transformations for Brans - Dicke theory
in a flat Friedmann - Lema\^{i}tre - Robertson - Walker (FLRW) mini-superspace case; 
in Sec.~\ref{gaugeFLRW} we show that this transformation is Hamiltonian canonical on the reduced phase space defined by gauge fixing the lapse function. \\
In Sec.~\ref{Sect:field} we generalize the previous considerations 
in the field theory case with the Arnowitt-Deser-Misner (ADM) formalism.
In this case too, see Sec.~\ref{gaugeADM}, we show that the transformation from JF to EF is Hamiltonian canonical 
on the reduced phase space. Here we gauge fixed both the lapse and the shifts functions. 
We argue on the very issue whether a Hamiltonian canonical transformation, between two systems, implies a physical equivalence and conclude in Sec.~\ref{Conclusions}.

\section{ Hamiltonian Weyl (conformal) transformations in FLRW Universe}
\label{Sect:finite}

In this Section and in the following we consider a flat mini-superspace FLRW model; the metric tensor is defined as:
\begin{equation}
g=-N^2(t) dt \otimes dt +a^2(t) dx^{i} \otimes dx^{i}
\label{FLRWpiatta}
\end{equation}
where $N(t)$ is the lapse function and $a(t)$ the scale factor. $N(t)$ is a real dynamical variable \cite{Christodoulakis:2013xha}, and not only a Lagrange multiplier \cite{thiemann2007}. 
Starting from the action \eqref{BDaction} we obtain 
the following Lagrangian in the JF \cite{Bonanno2017}:
\begin{equation}
\label{LagrJF}
{\mathcal{L}}_{FLRW}=-\frac{6a{\dot{a}}^2}{N}\phi-\frac{6a^2{\dot{a}}}{N}{\dot{\phi}}+\frac{\omega a^3}{N\phi}(\dot{\phi})^2-Na^3U(\phi)\,.  \end{equation}
Remember that in this FLRW case, assuming $\omega\neq-3/2$, the Brans-Dicke total Hamiltonian in the JF is \cite{Galaverni:2021jcy}:
\begin{eqnarray}
H_{\mathrm{T}}&=&N\bigg[-\frac{\omega{\pi}^{2}_{a}}{12a\phi (2\omega +3)}-\frac{ {\pi}_{a} \pi_{\phi}}{2 a^2 (2\omega +3)} 
+\frac{ \phi {\pi}^2_{\phi}}{2 a^3 (2\omega +3)}\nonumber\\
&& + a^3 U(\phi) \bigg]  + \lambda_N \pi_N \equiv N H + \lambda_N \pi_N\,,
\label{H:JF:neq}
\end{eqnarray}
we refer to Appendix \ref{AppA1a} for the conformal invariant case $\omega=-3/2$.

In this particular case the equations of motion are:
\begin{eqnarray}
\label{eq_neq32_N}
\dot{N}&\approx&\{N,H_T\}\approx \lambda_N \;,\\
\label{eq_neq32_pi_N}
\dot{\pi}_N&\approx&\{H,H_T\}\approx-H\,, \\
\label{eq_neq32_a}
\dot{a}&\approx&\{a,H_T\}\approx -\frac{N}{2 a (2\omega+3)}\left(\frac{\omega \pi_a}{3\phi}+\frac{\pi_\phi}{a}\right)\;,\\
\label{eq_neq32_pi_a}
\dot{\pi}_a &\approx&\{\pi_a,H_T\}\approx -\frac{N}{2a^2(2\omega+3)}\left(
\frac{\omega\pi_a^2}{6\phi}+\frac{2\pi_a\pi_\phi}{a}-\frac{3\phi\pi_\phi^2}{a^2}\right)\nonumber\\
&&-3 N a^2 U(\phi)\,,\\
\label{eq_neq32_phi}
\dot{\phi} &\approx&\{\phi,H_T\}\approx \frac{N}{2 a^2 (2\omega+3)}
\left(-\pi_a+\frac{2\phi \pi_\phi}{a}\right)\,,\\
\dot{\pi}_\phi &\approx&\{\pi_\phi,H_T\}\approx -\frac{N}{2 a (2\omega+3)}\left(\frac{\omega\pi_a^2}{6\phi^2}+\frac{\pi_\phi^2}{a^2}\right)
-Na^3\frac{dU}{d\phi}\,.\nonumber\\
\label{eq_neq32_pi_phi}
\end{eqnarray}

The Weyl (conformal) transformations \eqref{Weyltrans1} preserve the ADM structure:
\begin{equation}
\widetilde {g}=-\widetilde {N}^2(t) dt \otimes dt +\widetilde {a}^2(t) dx^{i} \otimes dx^{i}\,,
\label{FLRWpiattaEF}
\end{equation}
provided the following redefinitions of the lapse and scale factor (the scalar field is unchanged):
\begin{equation}
\label{JFEFtrans0}
{{\widetilde {N}}}=N (16\pi G\phi)^{\frac{1}{2}} \,  , \,  \,  \widetilde{a}=(16\pi G\phi)^{\frac{1}{2}}a\,  , \,  \,{{\widetilde {\phi}}}=\phi \,.
\end{equation}
If we apply this transformations to Eq.~\eqref{LagrJF}
we obtain the Lagrangian in the EF:
\begin{eqnarray}
\widetilde{\mathcal L}_{FLRW}&=&-\frac{1}{{\widetilde {N}(16\pi G\phi)}} \left[
6 \widetilde{a} \dot{\widetilde{a}}^2 \phi-\frac{(2\omega+3)\widetilde{a}^3\dot{{\phi}}^2}{2{\phi}}
\right]\nonumber\\
&&- {{\widetilde {N}}} \widetilde{a}^3 V(\phi)\,.
\end{eqnarray}
The total Hamiltonian in the EF is (see \cite{Galaverni:2021jcy}):
\begin{eqnarray}
\widetilde{H}_T &=&  \widetilde{N} \widetilde{a}^3  \left[
-\frac{2 \pi G \widetilde{\pi}_a^2}{3\widetilde{a}^4}
+\frac{8\pi G \widetilde{\pi}_\phi^2 \phi^2 }{(2\omega+3)\widetilde{a}^6} +V(\phi)
\right]\nonumber\\
&&+ \widetilde{\lambda}_N \widetilde{\pi}_N  
\equiv \widetilde{N}  \widetilde{H} +\widetilde{\lambda}_N \widetilde{\pi}_N \,,
 \label{H:EF:neq}
\end{eqnarray}
(we always assume $\omega\neq-3/2$, and refer to Appendix \ref{AppA1b} for the particular case $\omega=-3/2$).
%Performing the Hamiltonian analysis, 
The equations of motion in the EF are:
\begin{eqnarray}
\label{eq_neq32_JFEF_N}
\dot{\widetilde{N}}&\approx& \{\widetilde{N},\widetilde{H}_T\} \approx \widetilde{\lambda}_N \;,\\
\dot{\widetilde{\pi}}_N&\approx&\{\widetilde{\pi}_N,\widetilde{H}_T\} \approx -\widetilde{H}\,, \\
\dot{\widetilde{a}}&\approx&\{\widetilde{a},\widetilde{H}_T\} \approx  -\widetilde{N} \frac{4\pi G \widetilde{\pi}_a }{3 \widetilde{a}}\;,\\
\dot{\widetilde{\pi}}_a&\approx&\{\widetilde{\pi}_a,\widetilde{H}_T\}  \nonumber\\
&\approx&\widetilde{N} 
\left[-\frac{(2\pi G) \widetilde{\pi}_a^2 }{3 \widetilde{a}^2}
+\frac{3 (8\pi G)\widetilde{\pi}_\phi^2 \widetilde{\phi}^2 }{(2\omega+3)\widetilde{a}^4}
-3 \widetilde{a}^2 V(\widetilde{\phi}) \right]\,,\\
\dot{\widetilde{\phi}}&\approx&\{\widetilde{\phi},\widetilde{H}_T\} \approx  \widetilde{N} 
\frac{(16 \pi G )\widetilde{\pi}_\phi \widetilde{\phi}^2 }{(2\omega+3)\widetilde{a}^3}\,,\\
\dot{\widetilde{\pi}}_\phi&\approx&\{\widetilde{\pi}_\phi,\widetilde{H}_T\} \approx  -\widetilde{N} 
    \left[
     \frac{16 \pi G \widetilde{\pi}_\phi^2 \widetilde{\phi} }{(2\omega+3)\widetilde{a}^3}
    +\widetilde{a}^3 \frac{d V(\widetilde{\phi})}{d\widetilde{\phi}}
    \right]\,.
\label{eq_neq32_JFEF_pi_phi}
\end{eqnarray}

The remaining relations among the canonical variables in the EF and in the JF are \cite{Dyer,Deruelle2009,Gionti2021,Galaverni:2021xhd}:
\begin{eqnarray}
&& {\widetilde \pi}_{a}=\frac{{\pi}_{a}}{(16\pi G\phi)^{\frac{1}{2}}}   \, , \,\, {\widetilde \pi}_\phi=\frac{1}{\phi} ( \phi \pi_{\phi}-\frac{1}{2}a\pi_{a})\,, \nonumber\\
&& {\widetilde{\pi}}_N=\frac{\pi_N}{(16\pi G\phi)^{\frac{1}{2}}}\,.
\label{JFEFtrans}
\end{eqnarray}

\begin{figure}[ht]
\centering
\begin{tikzpicture}
  \matrix (m) [matrix of math nodes,row sep=4em,column sep=4em,minimum width=2em]
  {
    {H}_{T} & \widetilde{H}_{T} \\
     \mathrm{JF\, eqs.\,of\, mot.} & \mathrm{EF\, eqs.\,of\, mot.} \\};
  \path[stealth-stealth]
    (m-1-1) edge node [left] {} (m-2-1)
            edge [] node [above] {JF $\longleftrightarrow$ EF} (m-1-2)
    (m-1-2) edge node [right] {} (m-2-2);
\end{tikzpicture}
\caption{On the extended phase space we transform
${H}_{T}$ (the total Hamiltonian in the JF) 
into $\widetilde{H}_{T}$ (the total Hamiltonian in the EF) 
using the relations between variables and conjugate momenta in the two frames, see \eqref{JFEFtrans0} and \eqref{JFEFtrans}. 
It is not possible to pass from the equations of motion in the JF to the equations 
of motion in EF (and vice-versa) simply using the relations \eqref{JFEFtrans0} and \eqref{JFEFtrans}, 
for more details see \cite{Galaverni:2021jcy}. 
} 
\label{fig:M1}
\end{figure}
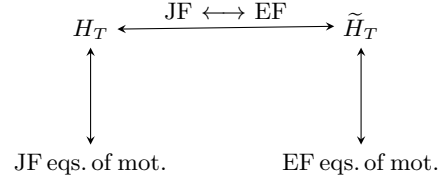

As we extensively discussed in \cite{Galaverni:2021jcy},
the transformations between JF and EF, see Eqs.  \eqref{JFEFtrans0} and \eqref{JFEFtrans}, are not Hamiltonian canonical 
transformations on the phase space defined 
by the canonical variables and their conjugate momenta (extended phase space).
The Poisson brackets among conjugate variables in the EF, 
expressed as function of the canonical variables in the JF, are:
\begin{eqnarray}
\label{cancond1}
\left\{{\widetilde {N}},{\widetilde{\pi}}_N  \right\}=1\,,\;
\left\{{\widetilde {a}},{\widetilde{\pi}}_a  \right\}=1\,,\;
\left\{{\phi},{\widetilde{\pi}}_\phi  \right\}=1\,.
\end{eqnarray}
Instead the Poisson brackets among non conjugate variables should be zero. This is certainly true for:
\begin{eqnarray}
&&\left\{{\widetilde {N}}, \widetilde{a}  \right\}=0\,,\;
\left\{{\widetilde {N}},{\widetilde{\pi}}_a  \right\}=0\,,\;
\left\{{\widetilde {N}},{\phi}  \right\}=0\,\,\nonumber\\
&&\left\{{\widetilde {a}}, \widetilde{\pi}_N  \right\}=0\,,\;
\left\{{\widetilde {a}},{\phi}  \right\}=0\,,\;
\left\{{\widetilde {a}},{\widetilde{\pi}}_\phi  \right\}=0\,,\label{cancond2}\\
&&\left\{{\phi}, \widetilde{\pi}_N  \right\}=0\,,\;
\left\{{\phi},{\widetilde{\pi}}_a  \right\}=0\,,\;
\left\{{\widetilde {\pi}}_N,{\widetilde{\pi}}_a  \right\}=0\,,\nonumber
\end{eqnarray}
but there are also Poisson brackets among non conjugate variables which does not vanishes:
\begin{eqnarray}
\label{non:canon}
\left\{{\widetilde {N}},{\widetilde{\pi}}_\phi  \right\}
&=&
\left\{N (16\pi G\phi)^{\frac{1}{2}}, \frac{1}{\phi} \, ( \phi \pi_{\phi}-\frac{1}{2}a\pi_{a}) \right\}\nonumber\\
&=&\frac{8\pi G N}{(16\pi G\phi)^{\frac{1}{2}}}\neq 0\,.
\end{eqnarray}
\begin{eqnarray}
\label{non:canon2}
\left\{{\widetilde {\pi}}_N,{\widetilde{\pi}}_\phi  \right\}
&=&
\left\{\frac{\pi_N}{(16\pi G\phi)^{\frac{1}{2}}}, \frac{1}{\phi} \, ( \phi \pi_{\phi}-\frac{1}{2}a\pi_{a}) \right\}\nonumber\\
&=&-\frac{8\pi G \pi_N}{(16\pi G\phi)^{\frac{3}{2}}}\neq 0\,.
\end{eqnarray}
Therefore the set of Weyl (conformal) transformations, see Eqs. \eqref{JFEFtrans0} and \eqref{JFEFtrans}, is not a Hamiltonian canonical map on the extended phase space.
We explicitly discussed this 
in \cite{Galaverni:2021jcy}, see also Fig.~\ref{fig:M1}. 
It is not possible to pass from the equations of motion in the EF to the equations 
of motion in JF (and viceversa) simply using the relations \eqref{JFEFtrans0} and \eqref{JFEFtrans}.

Once we apply these transformations on the EF equations of motion \eqref{eq_neq32_JFEF_N}-\eqref{eq_neq32_JFEF_pi_phi} we obtain the following:
\begin{eqnarray}
\label{eq_neq32_JFEF_inv_N}
\dot{N} &\approx& \frac{\widetilde{\lambda}_N}{(16\pi G\phi)^{\frac{1}{2}}}
-\frac{N^2}{2 a^2(2\omega+3)}\left(\frac{\pi_\phi}{a}-\frac{\pi_a}{2\phi}\right)\;, \\
\label{eq_neq32_JFEF_inv_pi_N}
\dot{\pi}_N &\approx& -H +\frac{N\pi_N}{2a^2(2\omega+3)}\left(\frac{\pi_\phi}{a}-\frac{\pi_a}{2\phi}\right)\,, \\
\label{eq_neq32_JFEF_inv_a}
\dot{a} &\approx& -\frac{N}{2 a (2\omega+3)}\left(\frac{\omega \pi_a}{3\phi}+\frac{\pi_\phi}{a}\right)\;,\\
\dot{\pi}_a &\approx& -\frac{N}{2a^2(2\omega+3)}\left(
\frac{\omega\pi_a^2}{6\phi}+\frac{2\pi_a\pi_\phi}{a}-\frac{3\phi\pi_\phi^2}{a^2}\right)\nonumber\\
&&-3 N a^2 U(\phi)\,,\label{eq_neq32_JFEF_inv_pi_a}\\
\label{eq_neq32_JFEF_inv_phi}
\dot{\phi} &\approx& \frac{N}{2 a^2 (2\omega+3)}\left(-\pi_a+\frac{2\phi \pi_\phi}{a}\right)\,,\\
\dot{\pi}_\phi &\approx& -\frac{N}{2 a (2\omega+3)} 
\left( \frac{\omega\pi_a^2}{4\phi^2}-\frac{7\pi_\phi^2}{2a^2}+\frac{\pi_a\pi_\phi}{2a\phi} \right)
\nonumber\\
&&-Na^3\frac{dU}{d\phi}+\frac{N a^3}{2\phi}U(\phi)\,.
\label{eq_neq32_JFEF_inv_pi_phi}
\end{eqnarray}
Looking at these transformed equations we note that: $(i)$ the equations for $\dot{a}\,,\dot{\pi}_a\,,\dot{\phi}$ coincide with JF Eqs. \eqref{eq_neq32_a}, \eqref{eq_neq32_pi_a}, \eqref{eq_neq32_phi};
$(ii)$ the equation for $\dot{\pi}_N$ corresponds to the original JF Eq. \eqref{eq_neq32_pi_N} only for $\pi_N=0$;
$(iii)$ using the Hamiltonian constrain the equation for $\dot{\pi}_\phi$ can be written as:
\begin{equation}
\dot{\pi}_\phi\approx-\frac{N}{2 a (2\omega+3)}\left(\frac{\omega\pi_a^2}{6\phi^2}+\frac{\pi_\phi^2}{a^2}\right)
-Na^3\frac{dU}{d\phi}+\frac{H}{2\phi}\,,
\label{eq_neq32_JFEF_inv_pi_phi:new}
\end{equation}
it corresponds to the original JF Eq. \eqref{eq_neq32_pi_phi} modulo the Hamiltonian constraint;
$(iv)$ the equation for $\dot{N}$ corresponds to the original JF Eq. \eqref{eq_neq32_N} only for a very particular choice of $\widetilde{\lambda}_N$:
\begin{equation} 
\widetilde{\lambda}_N= (16\pi G\phi)^{\frac{1}{2}}
\left[ \lambda_N 
+\frac{N^2}{2 a^2(2\omega+3)}\left(\frac{\pi_\phi}{a}-\frac{\pi_a}{2\phi}\right)\right]\,. \label{riucco}
\end{equation}

We have already seen in Eq. \eqref{JFEFtrans} that: 
$${\widetilde{\pi}}_N=\frac{{\pi}_N}{\left(16\pi G\phi\right)^{\frac{1}{2}}}\,,$$ 
if the extended Hamiltonian ${\widetilde{H}}_T$ is obtained applying the Hamiltonian transformations from the Jordan to the Einstein frame, then  $\widetilde{\lambda}_N= (16\pi G\phi)^{\frac{1}{2}}
 \lambda_N $. This formula is not equivalent to \eqref{riucco}, which is another way to check that the transformations from the Jordan to the Einstein frames are not Hamiltonian canonical, strictly speaking, on the extended phase space.

\section{ Gauge fixing in the FLRW case}
\label{gaugeFLRW}

We show here how it is possible to prove the canonicity of the transformation \eqref{JFEFtrans}
between JF and EF performing a gauge fixing on the lapse function $N$.
On the reduced phase space defined by this gauge fixing, the transformation is Hamiltonian canonical.
\subsection{Jordan frame}

We start with a particular gauge fixing condition for the lapse function in the JF:
\begin{equation}
\label{JFgaugef}
 N= c_0 \,, 
\end{equation}
here and in the following, $c_0=c_0(x)$ is an arbitrary function of the coordinates.
We implement this condition as a secondary constraint:
\begin{eqnarray}
\label{JFconstraint}
N-c_0\approx 0\,.
\end{eqnarray}

In order to avoid ambiguities, we remark, here and in the following, that \eqref{JFconstraint} does not mean that we are fixing the value of the lapse function. The lapse function continues to be an independent variable on the extended phase space.

Note that the Hamiltonian constraint remains a first class constraint even after the introduction 
of the new secondary constraint:
\begin{equation}
\{N-c_0,H\}\approx 0 \,.
\end{equation}
On the contrary, the first class constraint $\pi_N\approx 0$ 
becomes now a second class constraint after gauge fixing:
\begin{equation}
\{N-c_0,\pi_N\}\approx 1 \,.
\end{equation}
Therefore, in this particular case, we have two second class constraint:  
\begin{equation}
\chi_0\equiv N-c_0\,,\;\mbox{and}\;\chi_1\equiv \pi_N\,.    
\label{FLRW_JW_constr}
\end{equation}

Imposing the second class constraint to be preserved on the constraint surface, we get:
\begin{equation}
\dot{\chi}_0\approx\left\{N-c_0,H_T\right\}\approx0\,,
\end{equation}
which implies $\lambda_N\approx0$, and:
\begin{equation}
\dot{\chi}_1\approx\left\{\pi_N,H_T\right\}\approx0\,,    
\end{equation}
which is automatically preserved since $H\approx0$.

Following the Dirac procedure described in \cite{dirac1966,Henneaux:1992ig} 
we introduce the second class constraint matrix \cite{dirac1966}:
\begin{equation}
\label{Def:C}
C_{\alpha\beta}\equiv \{{\chi}_{\alpha},{\chi}_{\beta}\}\,.    
\end{equation}
and we obtain:
\begin{eqnarray}
C_{\alpha\beta}=\left(
\begin{array}{cc}
    0 & 1 \\
   -1 & 0
\end{array}
\right)\,,\;\mathrm{and}\; 
C_{\alpha\beta}^{-1}=\left(
\begin{array}{cc}
    0 & -1 \\
    1 & 0
\end{array}
\right)\,.
\end{eqnarray}
We are now ready to introduce the {\em Dirac brackets} (DB), they are defined starting from the Poisson brackets 
and the inverse of the second class constraint matrix \cite{dirac1966,Henneaux:1992ig}:
\begin{equation}
\left\{\,\cdot\,,\,\cdot\,\right\}_{DB}\equiv\left\{\,\cdot\,,\,\cdot\,\right\}-\left\{\,\cdot\,,\chi_\alpha\right\} C^{-1}_{\alpha\beta} \left\{\chi_\beta,\,\cdot\,\right\}\,.
\label{Diracbrackets1}
\end{equation}
Equations of motion are derived employing Dirac brackets. Afterward, we strongly impose the second class constraints and reduce the degrees of freedom.

The equations of motion for $a$, $\phi$,  $\pi_a$ and $\pi_\phi$, 
are:
\begin{eqnarray}
\dot{a}&\approx&\{a,H_T\}_{DB}\nonumber\\
 &\approx& -\frac{N}{2 a (2\omega+3)}\left(\frac{\omega \pi_a}{3\phi}+\frac{\pi_\phi}{a}\right)\;,
\label{adotJF}
\end{eqnarray}

\begin{eqnarray}
\dot{\pi}_a&\approx&\{{\pi}_a,H_T\}_{DB}\nonumber\\
&\approx& -\frac{N}{2a^2(2\omega+3)}\left(
\frac{\omega\pi_a^2}{6\phi}+\frac{2\pi_a\pi_\phi}{a}-\frac{3\phi\pi_\phi^2}{a^2}\right)\nonumber\\
&& -3 N a^2 U(\phi)\,,
\end{eqnarray}

\begin{eqnarray}
\dot{\phi}&\approx&\{\phi,H_T\}_{DB}\nonumber\\
&\approx&\frac{N}{2 a^2 (2\omega+3)}
\left(-\pi_a+\frac{2\phi \pi_\phi}{a}\right)\,,
\end{eqnarray}

\begin{eqnarray}
\dot{\pi}_\phi&\approx&\{\pi_{\phi},H_T\}_{DB}\nonumber\\
&\approx&-\frac{N}{2 a (2\omega+3)}\left(\frac{\omega\pi_a^2}{6\phi^2}+\frac{\pi_\phi^2}{a^2}\right)
-N a^3\frac{dU}{d\phi}\,.
\label{piphidotJF}
\end{eqnarray}
Strongly imposing the second class constraints $N=c_0$ and $\pi_N=0$, see Eqs. \eqref{FLRW_JW_constr}, we get the equations of motion on the reduces phase space:
\begin{eqnarray}
\label{adotJF2}
\dot{a}&\approx& -\frac{c_0}{2 a (2\omega+3)}\left(\frac{\omega \pi_a}{3\phi}+\frac{\pi_\phi}{a}\right)\;,\\
\dot{\pi}_a&\approx& -\frac{c_0}{2a^2(2\omega+3)}\left(
\frac{\omega\pi_a^2}{6\phi}+\frac{2\pi_a\pi_\phi}{a}-\frac{3\phi\pi_\phi^2}{a^2}\right)\nonumber\\
&& -3 c_0 a^2 U(\phi)\,,\\
\dot{\phi}&\approx&\frac{c_0}{2 a^2 (2\omega+3)}
\left(-\pi_a+\frac{2\phi \pi_\phi}{a}\right)\,,\\
\dot{\pi}_\phi&\approx&
-\frac{c_0}{2 a (2\omega+3)}\left(\frac{\omega\pi_a^2}{6\phi^2}+\frac{\pi_\phi^2}{a^2}\right)
-c_0 a^3\frac{dU}{d\phi}\,.
\label{piphidotJF2}
\end{eqnarray}

\subsection{Einstein frame}
\label{FLRW:EF:neq}

Gauge fixing in the JF \eqref{JFgaugef} implies the following gauge fixing in the EF:
\begin{equation}
\label{EFgaugef}
 \widetilde{N}= c_0  (16\pi G\phi)^{1\over 2}\,,
\end{equation}
this condition is implemented introducing an additional secondary constraint:
\begin{eqnarray}
\label{EFconstraint}
\widetilde{N}-c_0 (16\pi G\phi)^{1\over 2}  \approx 0\,.
\end{eqnarray}

Also in this frame the first class constraint $\widetilde{\pi}_N\approx 0$ now becomes a second class constraint 
after gauge fixing, because:
\begin{equation}
\{\widetilde{N}-c_0 (16\pi G\phi)^{1\over 2} ,\widetilde{\pi}_N\}\approx 1\,.    
\end{equation}

Here also the Poisson bracket between the Hamiltonian constraint defined in Eq.~\eqref{H:EF:neq} and 
the new gauge fixing constraint, see Eq.~\eqref{EFconstraint}, 
is different from zero:
\begin{equation}
\{\widetilde{N}-c_0 (16\pi G\phi)^{1\over 2},  \widetilde{H} \}=
-\frac{c_0 \left(16 \pi G \phi\right)^{3/2} \widetilde{\pi}_\phi}{2(2\omega+3) \widetilde{a}^3}\equiv -\eta
\neq 0
\label{nonfirstc}
\end{equation}
where we introduced $\eta$.

It looks that also $\widetilde{H}$ is now a second class constraint.
However, a linear combination of Hamiltonian constraint and the primary constraint defines a new Hamiltonian constraint 
\begin{equation}
\widetilde{H}^\prime\equiv\widetilde{H}+\eta\, \widetilde{\pi}_N\,,
\label{againfirst}
\end{equation}
which stays first class 
\begin{eqnarray}
 \{\widetilde{N}-c_0 (16\pi G\phi)^{1\over 2}, \widetilde{H}^\prime \} =-\frac{c_0^2 (16 \pi G)^2 \phi  \widetilde{\pi}_N}{4(2\omega+3) \widetilde{a}^3} \approx 0\,, 
\end{eqnarray}
The new total Hamiltonian is:
\begin{equation}
\widetilde{H}_T^\prime =  \widetilde{N}  \widetilde{H}^\prime +\widetilde{\lambda}_N \widetilde{\pi}_N \,.    
\end{equation}

We remain with two second class constraints:
\begin{equation}
\widetilde{\chi}_0\equiv \widetilde{N}-c_0(16\pi G\phi)^{1\over 2}\,,\;\mbox{and}\;
\widetilde{\chi}_1\equiv \widetilde{\pi}_N\,.    
\label{FLRW_EF_constr}
\end{equation}

These constraints are preserved if:
\begin{equation}
\dot{\widetilde{\chi}}_0\approx\left\{\widetilde{N}-c_0(16\pi G\phi)^{1\over 2},\widetilde{H}_T^\prime   \right\}\approx0\,,    \end{equation}
which implies $\widetilde{\lambda}_N\approx0$, and:
\begin{equation}
\dot{\widetilde{\chi}}_1\approx \left\{{\widetilde{\pi}}_N,\widetilde{H}_T^\prime\right\}\approx0\,,    
\end{equation}
which is automatically verified since $\widetilde{H}^\prime\approx0$.
The dynamics stays confined on the reduced phase space defined by the 
second class constraints.

The matrix of the (irreducible) second class constraints is:
\begin{eqnarray}
C_{\alpha\beta}\equiv\left(
\begin{array}{cc}
    0 & 1 \\
   -1 & 0
\end{array}
\right)\,,\;\mathrm{and}\,\,\,
C_{\alpha\beta}^{-1}\equiv\left(
\begin{array}{cc}
    0 & -1 \\
    1 & 0
\end{array}
\right)\,.
\end{eqnarray}

Now we have to evaluate only four equations of motion,
since, after gauge fixing \eqref{FLRW_EF_constr}, $\widetilde{N}$ and ${\widetilde{\pi}}_N$ are not anymore independent dynamical variables.
Always using Dirac brackets -defined in Eq.~\eqref{Diracbrackets1} - we get:
\begin{eqnarray}
\label{adotEF}
\dot{\widetilde{a}}&\approx&\left\{\widetilde{a},{\widetilde{H}}_T^\prime \right\}_{DB}\approx -\widetilde{N} \frac{4\pi G \widetilde{\pi}_a }{3 \widetilde{a}}
  \;,
\end{eqnarray}
\begin{eqnarray}
\dot{{\widetilde{\pi}}}_a&\approx&\left\{{\widetilde{\pi}_a},{\widetilde{H}}_T^\prime   \right\}_{DB}
 \nonumber\\
 &\approx&  
\widetilde{N} 
\left[-\frac{(2\pi G) \widetilde{\pi}_a^2 }{3 \widetilde{a}^2}
+\frac{3 (8\pi G)\widetilde{\pi}_\phi^2 \widetilde{\phi}^2 }{(2\omega+3)\widetilde{a}^4}
-3 \widetilde{a}^2 V(\phi) \right]\nonumber\\
&&-\frac{3 c_0  \left(16 \pi G \phi\right)^{3/2}  \widetilde{\pi}_\phi \widetilde{\pi}_N}{(2\omega+3)\widetilde{a}^4}
\,,
\end{eqnarray}

\begin{eqnarray}
\dot{\phi}&\approx&\left\{\phi,{\widetilde{H}}_T^\prime  \right\}_{DB}
 \nonumber\\
 &\approx&
 \widetilde{N} 
\frac{(16 \pi G )\widetilde{\pi}_\phi \phi^2 }{(2\omega+3)\widetilde{a}^3}
-\frac{c_0  \left(16 \pi G \phi\right)^{3/2}   \widetilde{\pi}_N}{(2\omega+3)\widetilde{a}^3}\,,
\end{eqnarray}

\begin{eqnarray}
\label{piphidotEF}
\dot{{\widetilde{\pi}}}_\phi&\approx&\left\{{\widetilde{\pi}_\phi},{\widetilde{H}}_T^\prime   \right\}_{DB}\nonumber\\
&\approx&\left\{{\widetilde{\pi}_\phi},{\widetilde{H}}_T^\prime  \right\}  \nonumber\\
 &&- \left\{{\widetilde{\pi}}_\phi, \widetilde{N}-c_0(16\pi G\phi)^{1\over 2}\right\}  C^{-1}_{01} \left\{\widetilde{\pi}_N,{\widetilde{H}}_T^\prime  \right\} \nonumber\\
&\approx& -\widetilde{N} 
\left[
 \frac{16 \pi G \widetilde{\pi}_\phi^2 \widetilde{\phi} }{(2\omega+3)\widetilde{a}^3}
+\widetilde{a}^3 \frac{d V(\phi)}{d\phi}
\right]\nonumber\\
&&+\frac{c_0  \widetilde{\pi}_\phi \widetilde{\pi}_N}{(2\omega+3)\widetilde{a}^3} \frac{3}{2} \left(16 \pi G \phi\right)^{1/2} 16 \pi G\nonumber\\
&&-\frac{16 \pi G c_0}{2 \left(16 \pi G \phi\right)^{1/2}} \widetilde{H}^\prime\,.
\end{eqnarray}
Strongly imposing the second class constraints $\widetilde{N}= c_0  (16\pi G\phi)^{1\over 2}  $ and $\widetilde{\pi}_N=0$, see Eqs. \eqref{FLRW_EF_constr}, and $\widetilde{H}^\prime=0$ we get the equations of motion on the reduced phase space:
\begin{eqnarray}
\label{adotEF2}
\dot{\widetilde{a}}&\approx&
-c_0  (16\pi G\phi)^{1\over 2}  \frac{4\pi G \widetilde{\pi}_a }{3 \widetilde{a}}\,,
\\
\dot{{\widetilde{\pi}}}_a&\approx& c_0 (16\pi G\phi)^{1\over 2}
\left[-\frac{(2\pi G) \widetilde{\pi}_a^2 }{3 \widetilde{a}^2}
+\frac{3 (8\pi G)\widetilde{\pi}_\phi^2 \widetilde{\phi}^2 }{(2\omega+3)\widetilde{a}^4}\right.\nonumber\\
&&\left.-3 \widetilde{a}^2 V(\phi) \right]\,,
\\
\dot{\phi}&\approx&c_0 (16\pi G\phi)^{1\over 2} \frac{(16 \pi G )\widetilde{\pi}_\phi \phi^2 }{(2\omega+3)\widetilde{a}^3}\,, \\
\dot{{\widetilde{\pi}}}_\phi&\approx&- c_0 (16\pi G\phi)^{1\over 2} 
\left[
 \frac{16 \pi G \widetilde{\pi}_\phi^2 \widetilde{\phi} }{(2\omega+3)\widetilde{a}^3}
+\widetilde{a}^3 \frac{d V(\phi)}{d\phi}\right]\,.    
\label{piphidotEF2}
\end{eqnarray}

Since the lapse and its conjugate momentum are not anymore dynamical variables,
the Hamiltonian transformation from the JF to the EF \eqref{JFEFtrans} is reduced to four independent dynamical variables, on which it is completely canonical (see \eqref{cancond1}-\eqref{non:canon2}).

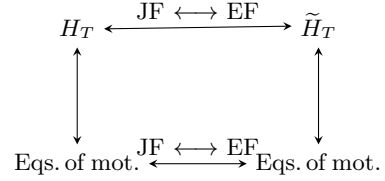
\begin{figure}[ht]
\centering
\begin{tikzpicture}
  \matrix (m) [matrix of math nodes,row sep=4em,column sep=4em,minimum width=2em]
  {
    {H}_{T} & \widetilde{H}_{T} \\
     \mathrm{Eqs.\,of\, mot.} & \mathrm{Eqs.\,of\, mot.} \\};
  \path[stealth-stealth]
    (m-1-1) edge node [left] {} (m-2-1)
            edge [] node [above] {JF $\longleftrightarrow$ EF} (m-1-2)
    (m-2-1.east|-m-2-2) edge node [below] {}
            node [above] {JF $\longleftrightarrow$ EF} (m-2-2)
    (m-1-2) edge node [right] {} (m-2-2);
\end{tikzpicture}
\caption{On the sub-manifold defined by the vanishing of the second class constraints the equations of motion in the EF \eqref{adotEF2}-\eqref{piphidotEF2} can be mapped into
the equations of motion in the JF \eqref{adotJF2}-\eqref{piphidotJF2}, and vice-versa,
using the transformations \eqref{JFEFtrans} together with the Hamiltonian constraint $H\approx0$  defined in Eq.~\eqref{H:JF:neq}.
This is a clear consequence of the canonicity of the Weyl (conformal)
transformation between JF and EF on the sub-manifold (reduced phase space) defined with the gauge fixing conditions.
} 
\label{fig:M2}
\end{figure}

A clear consequence of Hamiltonian canonical equivalence between JF and EF on the reduced phase space is the mathematical
equivalence between JF and EF also at the level of the
equations of motion, see Fig. \ref{fig:M2}.

\section{Hamiltonian Weyl (conformal) transformations in the general case}
\label{Sect:field}

The Arnowitt-Deser-Misner decomposition \cite{ADM} is based on the assumption that the topology of the Space-Time $(M,g)$ is $M={\mathbb{R}}\times \Sigma$ \cite{Esposito1992}, the metric tensor is defined as:
\begin{eqnarray}
g&=&-(N^{2}-N_{i}N^{i})dt \otimes dt +N_{i}(dx^{i} \otimes dt
+dt \otimes dx^{i})\nonumber\\
&&+h_{ij}dx^{i} \otimes dx^{j}
\label{ADMmetricJF}\,,
\end{eqnarray}
where $N\equiv N(t,x)$ is the lapse function, $N_i \equiv N_i(t,x)$ are the shifts functions and $h_{ij}\equiv h_{ij}(t,x)$ is the three-dimensional metric tensor on the three-dimensional surface $\Sigma$.

The ADM Lagrangian density $\mathcal{L}_{ADM}$ associated to the action \eqref{BDaction} is:
\begin{eqnarray}
\label{eq:Lagrangian2}
\mathcal{L}_{ADM}&=& {\sqrt{h}} \Bigg[N \phi\left( {}^{(3)}R+K_{ij}K^{ij}-K^2\right)\nonumber \\
&&-\frac{\omega}{N\phi}\left(N^2 h^{ij}D_i\phi D_j\phi- (\dot{\phi}-N^i D_i\phi)^2\right)  \\
&&+2K (\dot{\phi}-N^iD_i\phi )-NU(\phi)+2h^{ij}D_iN D_j\phi\Bigg]\;, \nonumber
\label{scompi}
\end{eqnarray}
where $K_{ij}$ is the extrinsic curvature defined as follows \cite{Esposito1992}:
\begin{equation}
{K}_{ij}=\frac{1}{2 N}
\left(-\frac{\partial h_{ij}}{\partial t} +D_{i}{N}_{j} +D_{j}{N}_{i}\right)\,.
\label{lextrins}
\end{equation}

The total Hamiltonian $H_T$, in the JF with $\omega\neq -\frac{3}{2}$ 
(see Appendix \ref{App2a} for the $\omega= -\frac{3}{2}$ case, and also \cite{Olmo})
is defined \cite{Gionti2021} as: 
\begin{equation}
H_{T}=\int d^{3}x \left(\lambda^{N} \pi_N + \lambda_{i}\pi^{i}+N{\mathcal{H}}+N_i{\mathcal{H}}^{i} \right)\,, 
\label{hamiltonianatot}
\end{equation}
where ${\mathcal{H}}$ is the Hamiltonian constraint \cite{Gionti2021}:
\begin{eqnarray}
{\mathcal{H}}&=&{\sqrt{h}}\Bigg\{ \left[-\phi\;  {}^{3}R+\frac{1}{\phi h}\left( \pi^{ij}\pi_{ij}-\frac{{\pi_h}^2}{2}\right)\right] \nonumber \\
&&+ \frac{\omega}{\phi}D_i\phi D^i\phi 
+2D^iD_i\phi                                     
 +U(\phi) \label{hamiltoeff1}  \\
 &&+\frac{1}{2 h\phi}\left(\frac{1}{3+2 \omega}\right)
 (\pi_h - \phi\pi_{\phi})^2\Bigg\} , 
  \nonumber
\end{eqnarray}
and we defined $\pi_h\equiv \pi^{ij} h_{ij}$.
${\mathcal{H}}^{i}$ are the momentum constraints 
\begin{equation}
{\mathcal {H}}^i= -2D_j\pi^{ji}+(D^i\phi) \pi_{\phi}\;. 
\label{momentumcons}
\end{equation}

The Weyl (conformal) transformations \eqref{Weyltrans1}  define the passage from the Jordan frame to the Einstein frame for Brans-Dicke theory. In general, one imposes that the ADM-metric in the Einstein frame is still
\begin{eqnarray}
\widetilde {g}&=&-({\widetilde N}^{2}-{\widetilde N}_{i}\widetilde{N}^{i})dt \otimes dt 
+\widetilde{N}_{i}(dx^{i} \otimes dt
+dt \otimes dx^{i})\nonumber\\
&&+\widetilde{h}_{ij}dx^{i} \otimes dx^{j}\,.
\label{EFmetricADM1}
\end{eqnarray}
Provided the following redefinitions of the variables:
\begin{eqnarray}
&&{{\widetilde {N}}}=N (16\pi G\phi)^{1\over 2} \,,\,\, {{\widetilde {N_i}}}=N_i (16\pi G\phi)\, , \nonumber\\
&&\, \widetilde{h}_{ij}=(16\pi G\phi)h_{ij} \, ,\,\, {{\widetilde {\phi}}}=\phi \, ,
\label{JFEFtrans2a}
\end{eqnarray}
we easily obtain the ADM Lagrangian density $\widetilde{\mathcal{L}}_{ADM}$ in the EF starting from \eqref{eq:Lagrangian2}:
\begin{eqnarray}
\label{eq:Lagrangian2EF}
\widetilde{\mathcal{L}}_{ADM}&=&
\frac{{\sqrt{\widetilde{h}}}}{16\pi G}
 \Bigg[\widetilde{N} \left( {}^{(3)}\widetilde{R}+\widetilde{K}_{ij}\widetilde{K}^{ij}-\widetilde{K}^2\right)\nonumber \\
&&-\frac{2\omega+3}{2 \widetilde{N} \phi^2}\left(\widetilde{N}^2 \widetilde{h}^{ij}\widetilde{D}_i\phi \widetilde{D}_j\phi- (\dot{\phi}-\widetilde{N}^i \widetilde{D}_i\phi)^2\right)\nonumber  \\
&&-16 \pi G\widetilde{N} V(\phi) \Bigg]\,.
\end{eqnarray}

The total Hamiltonian ${\widetilde{H}}_{T}$ in the EF, see \cite{Gionti2021}, is:
\begin{equation}
{\widetilde {H}}_{T}=\int d^{3}x \left({\widetilde{\lambda}}^{N} {\widetilde{\pi}}_N + {\widetilde{\lambda}}_{i}{\widetilde{\pi}}^{i}+{\widetilde{N}}{\widetilde{{\mathcal{H}}}}+{\widetilde{N}}_i{\widetilde{{\mathcal{H}}}}^{i} \right)\;\,, 
\label{hamiltonianatotEF}
\end{equation}
where 
\begin{eqnarray}
&&{\widetilde{\mathcal{H}}}=\frac{\sqrt{{\widetilde h}}}{16\pi G}\Bigg[ -{}^{3}{\widetilde R}+\frac{(16\pi G)^2}{\widetilde h}\left( {\widetilde \pi}^{ij}{\widetilde \pi}_{ij}-\frac{{{\widetilde \pi}_h}^2}{2}\right) \nonumber \\
&&+ \frac{(\omega +\frac{3}{2})}{{\phi}^2} {\widetilde D}_i\phi {\widetilde D}^i\phi 
+\frac{64(\pi G)^{2}{\phi}^{2}}{\widetilde{h}(\omega + \frac{3}{2})} {\widetilde \pi}_{\phi}^{2}\Bigg]
+\sqrt{{\widetilde h}}V(\phi)\,, \nonumber\\
\label{hamiltonianconsEF}
\end{eqnarray}
and 
\begin{equation}
{\widetilde{\mathcal{H}}}^{i}=-2{\widetilde D}_j{\widetilde \pi}^{ji}+{\widetilde D}^i\phi {\widetilde \pi}_{\phi}\,,
\label{momentaconstraint}
\end{equation}
(see Appendix \ref{App2b} for the $\omega= -\frac{3}{2}$ case).

We have already studied the Hamiltonian analysis of Branse-Dicke theory \cite{Gionti2021,Galaverni:2021xhd}, see also \cite{Olmo,Gielen,floreaniniJackiw,costagirotti,faddeevJackiw,Kiefer2017,Barvinsky2019}. There it was shown that the Weyl (conformal) transformations from the Jordan to the Einstein frame \eqref{Weyltrans1}
in this ADM case for momenta are:
\begin{eqnarray}
&& {\widetilde{\pi}}_N=\frac{\pi_N}{(16\pi G\phi)^{1\over 2}}  \,,\,\, {\widetilde{\pi}}^i=\frac{\pi^i}{(16\pi G\phi)}  \,,\nonumber\\
&&
{\widetilde{\pi}}^{ij}=\frac{\pi^{ij}}{(16\pi G\phi)^{1\over 2}}  \,,\,\, {\widetilde \pi}_\phi=\frac{1}{\phi} \, ( \phi \pi_{\phi}-\pi_{h})\,,
\label{JFEFtrans2b}
\end{eqnarray}
where $\pi_{h}\equiv\pi^{ij} h_{ij}$.

The transformations \eqref{JFEFtrans2a}-\eqref{JFEFtrans2b} are not Hamiltonian canonical transformations on the extended phase space \cite{Gionti2021,Galaverni:2021xhd}.
In fact, the Poisson brackets among the conjugate variables in the Einstein frame, expressed as function of the canonical variables in the Jordan frame,  
are equal to the delta function:
\begin{eqnarray}
\left\{\widetilde {N}(x),\widetilde{\pi}_N (x') \right\}&=& \delta^{(3)}(x-x')\,, \nonumber\\
\left\{\widetilde {N}_{i}(x),\widetilde{\pi}^{j} (x') \right\}&=& \delta_{i}^{j}\delta^{(3)}(x-x')\,, \nonumber\\
\left\{\widetilde {h}_{ij}(x),\widetilde{\pi}^{kl} (x') \right\}&=&\delta_{i}^{k}\delta_{j}^{l}\delta^{(3)}(x-x')\,, \nonumber \\
\left\{\phi(x),\widetilde{\pi}_{\phi} (x') \right\}&=& \delta^{(3)}(x-x')\, ,
\label{coniugate}
\end{eqnarray}
while the Poisson brackets among non-conjugate variables are, in general, zero, e.g.\cite{Deruelle2009}:
\begin{eqnarray}
&&\left\{\widetilde {N}(x),\widetilde{\pi}^{j} (x') \right\}=\left\{\widetilde {N}_i(x),\widetilde{\pi}_{N} (x') \right\}=\left\{\widetilde {\pi}^{i}(x),  \widetilde{\pi}_{\phi}(x')\right\} \nonumber\\
&&=\left\{\widetilde {N}(x),\widetilde{N}_{j} (x') \right\}=\left\{\widetilde {\pi}_N(x),\widetilde{\pi}^{j} (x')\right\}\nonumber \\ 
&&=\left\{\widetilde {N}(x),\widetilde{h}^{ij} (x')\right\}=
\left\{\widetilde {N}(x),\widetilde{\pi}^{ij} (x') \right\}=\left\{\widetilde {\pi}^i(x),\phi(x')\right\}
\nonumber \\
&&=\left\{\widetilde {N}(x),\phi (x') \right\}
=\left\{\widetilde {N}_i(x),\widetilde{h}^{ij} (x')\right\}=\left\{\widetilde {N}_h(x),\widetilde{\pi}^{ij} (x') \right\}
\nonumber\\
&&=\left\{\widetilde {N}_i(x),\phi (x') \right\}
=\left\{\widetilde {\pi}_N(x),\widetilde{h}_{ij} (x')\right\}=\left\{\widetilde {\pi}_N(x),\widetilde{\pi}^{ij} (x')\right\}
\nonumber \\
&&
=\left\{\widetilde {\pi}^i(x),\widetilde{\pi}^{ij} (x')\right\}
=\left\{\widetilde {\pi}^i(x),\widetilde{h}_{ij} (x')\right\}=\left\{\widetilde {\pi}_N(x),  \phi(x')\right\}
\nonumber \\
&&=\left\{\widetilde {h}_{ij}(x), \phi(x') \right\}
=\left\{\widetilde {h}_{ij}(x),\widetilde{\pi}_{\phi} (x') \right\}=\left\{\widetilde {\pi}^{ij}(x),  \phi(x')\right\}
\nonumber\\
&&=\left\{\widetilde {\pi}^{ij}(x),  \widetilde{\pi}_{\phi}(x')\right\}=0\,,
\label{quellechecommutano}
\end{eqnarray}
but in the following cases,
\begin{eqnarray}
\{{\widetilde N}(x),{\widetilde \pi}_{\phi}(x') \}&=&\frac{8\pi GN(x)\delta^{(3)}(x-x')}{{\sqrt{16 \pi G\phi(x)}}} \neq 0\,,\nonumber\\
\{{\widetilde N}_i(x),{\widetilde \pi}_{\phi}(x') \}&=&16 \pi G N_i(x)\delta^{(3)}(x-x')\neq 0\,, \nonumber\\
\{{\widetilde \pi}_N (x),{\widetilde \pi}_{\phi}(x') \}&=&-\frac{8\pi G\pi_N(x)\delta^{(3)}(x-x')}{\sqrt{\left(16 \pi G\phi(x)\right)^3}} \neq 0\,, \nonumber \\
\{{\widetilde \pi}^i(x),{\widetilde \pi}_{\phi}(x') \}&=& -  \frac{\pi ^i (x)}{(16\pi G \phi^{2})}\delta^{(3)}(x-x')\neq 0\,,
\label{noncanonicalcond2}
\end{eqnarray}
the Poisson brackets among non-conjugate variables are not zero.

\section{Gauge fixing in the general case}
\label{gaugeADM}

The strategy suggested by Eqs. \eqref{noncanonicalcond2},
as we proceeded in Sec.~\ref{gaugeFLRW}, is to gauge fix the lapse  $N$ 
and also the shifts $N_{i}$ field-variables and implement these gauge fixing conditions
as a secondary constraints. These secondary constraints make the momenta conjugated to the lapse $\pi_N$ 
and to the shifts $\pi_i$ second class constraints. 
Lapse and shifts continue to be canonical variables as they appear in the definition of the Poisson brackets.
After having defined the Dirac's brackets associated to these second class constraints \cite{dirac1966}, we can solve explicitly the second class constraints and express the lapse $N$, the shifts $N_i$ and their conjugated momenta $\pi_N$ and $\pi_i$ as functions of the other field variables.
In this way, we reduced the degrees of freedom of our system to a reduced phase space. 
On this phase space, we will explicitly show that the Hamiltonian conformal (Weyl) transformations from the Jordan to the Einstein frames are Hamiltonian canonical transformations.      

\subsection{Jordan frame}
\label{JoFr}
 We start with the gauge-fixing conditions for a Brans-Dicke theory \cite{Gionti2021} in the JF
\begin{equation}
N=c_0\;,\;\; N_{i}=c_i\,, 
\label{JFimpongo}
\end{equation}
being $c_0=c_0(x)$ and $c_i=c_i(x)$ arbitrary functions.
Implementing them  as a secondary constraints,
\begin{equation}
N-c_0 \approx 0\;,\;\; N_i -c_i \approx 0\;\;,
\label{secondario}
\end{equation}
we, immediately, notice that the primary first class constraints \cite{Gionti2021} 
\begin{equation}
\pi_{N}\approx 0 \;,\;\; \pi^{i}\approx 0 \;,
\label{primary-first}
\end{equation}
becomes second class constraints. In fact, we have 
\begin{eqnarray}
    \{ N(x)-c_0,\pi_N(x')\}\approx\delta^{(3)}(x-x')\,,\nonumber\\ 
    \{N_i (x) - c_i, \pi^{j}(x')\}=\delta^{j}_{i}\,\delta^{(3)}(x-x')\,.
    \label{secondary}
\end{eqnarray}

The Poisson brackets of the secondary gauge-fixing constraints with the secondary first class constraints in the JF \cite{Gionti2021} are:
\begin{eqnarray}
&&\{N-c_0, {\mathcal{H}}\}\approx 0 \;\; ,\;\; \{N_0 - c_0, {\mathcal{H}}^i\}=0\, ,\\ 
&&\{N_i - c_i, {\mathcal{H}}\}\approx 0 \;\; ,\;\; \{N_i - c_i, {\mathcal{H}}^i\}=0.
\label{gaugecommuto}
\end{eqnarray}

This is enough to check that the Hamiltonian constraint ${\mathcal{H}}$ and the momentum constraints ${\mathcal{H}}^i$ \cite{Gionti2021} remain first class constraints. 
Summarising the second constraints are:
\begin{eqnarray}
&&\chi_0 \equiv N -c_0\,\,,\,\, \chi_{i}\equiv N_i -c_i\,\,,\,\, \nonumber\\
&&\chi_4 \equiv \pi_N\,\,,\,\, \chi_{i+4} \equiv \pi_i\,\,.
\label{second}
\end{eqnarray}
These constraints are preserved if:
\begin{equation}
\dot{\chi}_0\approx \{N-c_0,H_T \}\approx0\,,
\end{equation}
which implies $\lambda^N(x)\approx0$, and if:
\begin{equation}
\dot{\chi}_i\approx \{N_i-c_i,H_T \}\approx0\,,
\end{equation}
which implies $\lambda_i(x)\approx0$.
The other two constraints: 
\begin{eqnarray}
\dot{\chi}_4&\approx& \{\pi_N,H_T \}\approx0\,,\\
\dot{\chi}_{i+4}&\approx& \{\pi^i,H_T \}\approx0\,,
\end{eqnarray}
are automatically preserved since ${\mathcal{H}}^i\approx0$ and ${\mathcal{H}}\approx0$.

The next step of the Dirac's procedure for constrained systems \cite{dirac1966} is to define the Dirac's brackets  using second class constraints \eqref{second};
 the inverse of second class constraint matrix, defined in Eq.~\eqref{Def:C}, is:
\begin{equation}
C^{-1}_{\alpha\beta}\equiv 
 \left(
     \begin{array}{c c c c}
      0 & 0 & -1 & 0 \\
    0 & 0 &  0 & -\mathbb{I} \\
    1 & 0 & 0  &  0   \\
    0 & \mathbb{I} & 0 & 0
      \end{array}
    \right)\,,
\label{matrix}
\end{equation}
where $\mathbb{I}$ is a $3 \times 3$ identity matrix. 

The Dirac's brackets (DB) are defined, through the Poisson brackets, following Eq.~\eqref{Diracbrackets1}.
Then, we derive the equations of motion using these brackets and afterwards we strongly impose the second class constraints \cite{dirac1966}.
It is very straightforward to check that, due to the particular structure of the Dirac's brackets \eqref{Diracbrackets1} the constraint algebra, among the secondary first class constraints, does not change if we replace the Poisson brackets with the Dirac's ones: 
\begin{eqnarray}
\{{\mathcal{H}}_i(x) ,{\mathcal{H}}_j(x')\}_{DB}&=&\{{\mathcal{H}}_i(x) ,{\mathcal{H}}_j(x')\}\,\,, \nonumber\\
\{{\mathcal{H}}_i(x) ,{\mathcal{H}}(x')\}_{DB}&=&\{{\mathcal{H}}_i(x) ,{\mathcal{H}}(x')\}\,\, , \nonumber\\
\{{\mathcal{H}}(x) ,{\mathcal{H}}(x')\}_{DB}&=&\{{\mathcal{H}}(x) ,{\mathcal{H}}(x')\}\,\, .
\label{equivo}
\end{eqnarray}

Now we are ready to write the equations of motion using Dirac's brackets. As in the finite dimensional case, ($N$, $N_i$, $\pi_N$,$\pi^i$) are not anymore dynamical variables, as consequences of the gauge fixing. The dynamics stays confined on the surface defined by second class constraints \eqref{secondario}-\eqref{primary-first} using Dirac's Brackets.

The equations of motion, using Dirac's brackets, for the other dynamical variables are the same, as it is easily to see, as those obtained using Poisson brackets. Starting with $h_{ij}$, we have
\begin{eqnarray}
\dot{h}_{ij}&\approx&\{h_{ij}, H_T\}_{DB}\nonumber\\
 &\approx&D_iN_j+D_jN_i+ \frac{2N}{\phi \sqrt{h}} \left(\pi_{ij}-\frac{\pi_h}{2} h_{ij} \right)\nonumber\\
&&+\frac{N}{\phi \sqrt{h}}\frac{(\pi_h -\phi \pi_{\phi})}{(2\omega + 3)}h_{ij}\,.
\label{metrico}
\end{eqnarray}

The equations of motion for the relative momenta $\pi^{ij}$ are 
\begin{eqnarray}
{\dot{\pi}}^{ij}&\approx&\{{\dot{\pi}}_{ij}, H_T\}_{DB}\nonumber\\
&\approx&-N\sqrt{h}\phi \left( {}^{(3)}R^{ij}- \frac{h^{ij}}{2}\, {}^{(3)}R \right) \nonumber\\
&&+\sqrt{h}\left( D^i D^j-h^{ij} D^k D_k\right)(N\phi) \nonumber \\
&&+\frac{Nh^{ij}}{2 \phi\sqrt{h}}\left( \pi^{ij}\pi_{ij}-\frac{{\pi}^{2}_{h}}{2}\right)+\frac{N}{\phi\sqrt{h}}\pi_h \pi^{ij}\nonumber\\
&&-\sqrt{h}\frac{N\omega}{\phi}D^i \phi D^j\phi-\sqrt{h}\frac{N\omega}{2\phi}h^{ij}D_k\phi D^k\phi \nonumber \\
&&+\sqrt{h}h^{ij}D_k N D^k \phi \nonumber \\
&&+ \sqrt{h}\left(D^i N D^j \phi + D^j N D^i \phi \right) -\frac{\sqrt{h}}{2} h^{ij} N U(\phi) \nonumber\\
&&+\frac{h^{ij}}{4 \sqrt{h}\phi}\left(\frac{N}{2\omega+3} \right)\left(\pi_{h}-\phi \pi_{\phi}\right)^{2}\nonumber\\
&&-\frac{\pi^{ij}}{\sqrt{h}\phi}\left(\frac{N}{2\omega+3} \right)\left(\pi_{h}-\phi \pi_{\phi}\right) \nonumber\\
&&+D_k \left(N^{k}\pi^{ij} \right)-\frac{2N}{\sqrt{h}\phi}\pi^{iq}\pi_{q}^{\,\,j} \nonumber\\
&&-(D_kN^{i})\pi^{kj}-(D_kN^{j})\pi^{ki}\, .
\label{pilunga}
\end{eqnarray}

The equation of motion for $\phi$ turns to be
\begin{eqnarray}
\dot \phi &&\approx \{\phi, H_T\}_{DB}\nonumber\\
&&\approx N^i D_i \phi -\frac{N}{\sqrt{h}(2\omega +3)}\left(\pi_h-\phi \pi_{\phi} \right)\,.
\label{phiequation}
\end{eqnarray}

Finally, the equation of motion for $\pi_{\phi}$ is 
\begin{eqnarray}
{\dot{\pi}}_{\phi}&\approx&  \{{\pi}_{\phi}, H_T\}_{DB}\nonumber\\
&\approx& \sqrt{h}N\, {}^{(3)}R +\frac{N}{{\phi}^2 \sqrt{h}}
\left(\pi^{ij}\pi_{ij}-\frac{{\pi_h}^2}{2} \right) \nonumber \\
&&+\frac{\sqrt{h}N\omega}{\phi^2}D_i\phi D^i\phi + 2\sqrt{h}D_i \left(\frac{N\omega}{\phi} D^{i}\phi\right) \nonumber\\
&&-2\sqrt{h}(D^iD_i)(N)-N\sqrt{h}\frac{dU}{d\phi}+D_i(N^i\pi_{\phi}) \nonumber\\
&&+\frac{N(\pi_h -\phi \pi_{\phi})^2}{2\sqrt{h}\phi^{2}(2\omega +3)}+
\frac{N(\pi_h -\phi \pi_{\phi})\pi_{\phi}}{\sqrt{h}\phi(2\omega +3)}.
\label{equopiphi}
\end{eqnarray}

Once we have calculated the equations of motion using Dirac's brackets, we impose strongly the second class constraints 
\eqref{secondario}-\eqref{primary-first} 
and implement them into Eqs. \eqref{metrico}-\eqref{equopiphi} to get the equations of motion defined on submanifold of the second class constraints:
\begin{eqnarray}
\dot{h}_{ij}\approx \frac{2c_0}{\phi \sqrt{h}} \left(\pi_{ij}-\frac{\pi_h}{2} h_{ij} \right)
+\frac{c_0}{\phi \sqrt{h}}\frac{(\pi_h -\phi \pi_{\phi})}{(2\omega + 3)}h_{ij}\,,
\label{metrico2}
\end{eqnarray}
\begin{eqnarray}
{\dot{\pi}}^{ij}&\approx&-c_0 \sqrt{h}\phi \left( {}^{(3)}R^{ij}- \frac{h^{ij}}{2}\, {}^{(3)}R \right) \nonumber\\
&&+c_0 \sqrt{h}\left( D^i D^j-h^{ij} D^k D_k\right)(\phi) \nonumber \\
&&+\frac{c_0 h^{ij}}{2 \phi\sqrt{h}}\left( \pi^{ij}\pi_{ij}-\frac{{\pi}^{2}_{h}}{2}\right)+\frac{c_0}{\phi\sqrt{h}}\pi_h \pi^{ij}\nonumber\\
&&-\sqrt{h}\frac{c_0 \omega}{\phi}D^i \phi D^j\phi-\sqrt{h}\frac{c_0 \omega}{2\phi}h^{ij}D_k\phi D^k\phi \nonumber \\
&& -\frac{\sqrt{h}}{2} h^{ij} c_0 U(\phi) \nonumber\\
&&+\frac{h^{ij}}{4 \sqrt{h}\phi}\left(\frac{c_0}{2\omega+3} \right)\left(\pi_{h}-\phi \pi_{\phi}\right)^{2}\nonumber\\
&&-\frac{\pi^{ij}}{\sqrt{h}\phi}\left(\frac{c_0}{2\omega+3} \right)\left(\pi_{h}-\phi \pi_{\phi}\right) \nonumber\\
&&+c^k D_k \left(\pi^{ij} \right)-\frac{2c_0}{\sqrt{h}\phi}\pi^{iq}\pi_{q}^{\,\,j} \,,
\label{pilunga2}
\end{eqnarray}
\begin{eqnarray}
\dot \phi \approx c^i D_i \phi -\frac{c_0}{\sqrt{h}(2\omega +3)}\left(\pi_h-\phi \pi_{\phi} \right)\,,
\label{phiequation2}
\end{eqnarray}
\begin{eqnarray}
{\dot{\pi}}_{\phi}&\approx& c_0\sqrt{h}\, {}^{(3)}R +\frac{c_0}{{\phi}^2 \sqrt{h}}
\left(\pi^{ij}\pi_{ij}-\frac{{\pi_h}^2}{2} \right) \nonumber \\
&&+c_0\frac{\sqrt{h}\omega}{\phi^2}D_i\phi D^i\phi + 2c_0\sqrt{h}D_i \left(\frac{\omega}{\phi} D^{i}\phi\right) \nonumber\\
&&-c_0\sqrt{h}\frac{dU}{d\phi}+c^{i}D_i(\pi_{\phi}) \nonumber\\
&&+\frac{c_0(\pi_h -\phi \pi_{\phi})^2}{2\sqrt{h}\phi^{2}(2\omega +3)}+
\frac{c_0(\pi_h -\phi \pi_{\phi})\pi_{\phi}}{\sqrt{h}\phi(2\omega +3)}.
\label{equopiphi2}
\end{eqnarray}

\subsection{Einstein frame}
\label{EF1}

The analogous of the gauge-fixing conditions \eqref{JFimpongo} in the EF are obtained using the transformation \eqref{JFEFtrans2a}:
\begin{equation}
\widetilde{N}=c_0\left( 16\pi G \phi\right)^{1\over 2}\;\;,\;\; {\widetilde{N}}_{i}=c_i\left( 16\pi G \phi\right)\,.
\label{EFimpongo1}
\end{equation}
First notice that the gauge conditions in the Jordan frame \eqref{JFimpongo} fix the gauge condition in the Einstein frame. These, as above, are implemented as secondary constraints 
\begin{equation}
\widetilde{N}-c_0\left( 16\pi G \phi\right)^{1\over 2} \approx 0\;\;,\;\; {\widetilde{N}}_{i}-c_i\left( 16\pi G \phi\right)\, \approx 0.
\label{EFimpongo}
\end{equation}

As in the Jordan frame, see Eqs. \eqref{secondary}, the primary first class constraints  
becomes second class constraints \cite{Gionti2021}:
\begin{eqnarray}
    &&\{\widetilde{N}(x)-c_0\left( 16\pi G \phi\right)^{1\over 2},\widetilde{\pi}_N (x')\}\approx \delta^{(3)}(x-x') \;,\; \nonumber\\
     &&\{{\widetilde{N}}_i (x) - c_i\left( 16\pi G \phi (x) \right), \widetilde{\pi}^{j}(x')\}\approx \delta^{j}_{i} \delta^{(3)}(x-x').\nonumber\\
    \label{secondaryEF}
\end{eqnarray}
The Poisson brackets of the secondary gauge-fixing constraints with the secondary first class constraints in the JF, case $\omega \neq -\frac{3}{2}$, are 
\begin{eqnarray}
&&\{\widetilde{N}(x)-c_0(16\pi G \phi(x))^{\frac{1}{2}},{\widetilde{\mathcal{H}}}(x')\}\nonumber\\
&&\approx -\frac{(16\pi G\phi(x))^{\frac{3}{2}}c_0\widetilde{\pi}_{\phi}(x)
\delta^{(3)}(x-x')}{4\sqrt{\widetilde{h}(x)}(\omega+\frac{3}{2})},
\label{primononcomm}
\end{eqnarray}
\begin{eqnarray}
&&\{\widetilde{N}(x)-c_0(16\pi G \phi(x))^{\frac{1}{2}},{\widetilde{\mathcal{H}}^{i}}(x')\} \nonumber\\
&&\approx -\frac{8\pi G\,\phi(x)\, c_0\, D^{i}\phi(x)
\delta^{(3)}(x-x')}{(16\pi G \phi(x))^{\frac{1}{2}}},
\label{secondononcomm}
\end{eqnarray}
\begin{eqnarray}
&&\{\widetilde{N}_i(x)-c_i(16\pi G \phi(x)),{\widetilde{\mathcal{H}}(x')}\} \nonumber\\
&&\approx-\frac{128c_i\widetilde{\pi}_{\phi}(x)\delta^{(3)}(x-x')(\pi G\,\phi(x))^2\,}{\sqrt{\widetilde{h}(x)}(\omega+\frac{3}{2})},
\label{terzononcomm}
\end{eqnarray}
\begin{eqnarray}
&&\{\widetilde{N}_i(x)-c_i(16\pi G \phi(x)),{\widetilde{\mathcal{H}}^{j}(x')}\} \nonumber\\
&&\approx-(16\pi G)c_i D^{j}\phi(x)\delta^{(3)}(x-x')\,.
\label{quartononcomm}
\end{eqnarray}

 It looks that the Hamiltonian constraint $\mathcal{H}$ and the momentum constraints ${\mathcal{H}}_i$ are made second class by these secondary gauge constraints, a phenomenon already observed in \eqref{nonfirstc}. Since we have implemented gauge condition only to reduce the redundant variables of the momenta associated to the lapse $N$ and the shifts $N_i$, we expect, as in Eq.~\eqref{againfirst}, that some linear combinations of the Dirac's constraints with a  suitable re-definitions of the Hamiltonian and momentum constraints will keep them still first class. 
 Elementary considerations, as in Section \ref{gaugeFLRW}, suggest to re-define the Hamiltonian and momentum constraints as 
 \begin{eqnarray}
{\widetilde{\mathcal{H}}}'&\equiv&\widetilde{\mathcal{H}} +\eta_N {\widetilde{\pi}}_N+\gamma_i {\widetilde{\pi}}^{i}\,, \nonumber\\
{{{\widetilde{\mathcal{H}}}^{i}}}{}'&\equiv&{\widetilde{\mathcal{H}}}^i +\eta^i {\widetilde{\pi}}_N+{\rho}_{k}^{\,\,i} {\widetilde{\pi}}^{k}\,\,,
\label{ridefinio}
\end{eqnarray}
where: 
\begin{equation}
\eta_N \equiv  \frac{(16\pi G\phi(x))^{\frac{3}{2}}c_0\widetilde{\pi}_{\phi}(x)
}{4\sqrt{\widetilde{h}(x)}(\omega+\frac{3}{2})},
\label{etaN}
\end{equation}
\begin{equation}
\eta^i \equiv \frac{8\pi G\,\phi(x)\, c_0\, D^{i}\phi(x)}
{(16\pi G \phi(x))^{\frac{1}{2}}},
\label{etai}
\end{equation}
\begin{equation}
\gamma_i \equiv \frac{128c_i\widetilde{\pi}_{\phi}(x)(\pi G\,\phi(x))^2\,}{\sqrt{\widetilde{h}(x)}(\omega+\frac{3}{2})},
\label{gammai}
\end{equation}
\begin{equation}
\rho_{k}^{\,\,i} \equiv (16\pi G)c_k D^{i}\phi(x)\,.
\label{rhoki}
\end{equation}

It is straightforward to check that now ${\widetilde{\mathcal{H}}}'$ and ${{{\widetilde{\mathcal{H}}}^{i}}}{}'$ are first class constraints, while $\pi_N $, $\pi^i$, $\widetilde{N}-c_0\left( 16\pi G \phi\right)^{1\over 2}$ and ${\widetilde{N}}_{i}-c_i\left( 16\pi G \phi\right)$
are second class constraints. Therefore, if we re-define, as in \eqref{second}, the second class constraints 
\begin{eqnarray}
&&{\widetilde{\chi}}_0 \equiv\widetilde{N}-c_0\left( 16\pi G \phi\right)^{1\over 2}\,\,,\,\, {\widetilde{\chi_{i}}}\equiv{\widetilde{N}}_{i}-c_i\left( 16\pi G \phi\right)\,\,,\,\, \nonumber\\
&&{\widetilde{\chi}}_4 \equiv {\widetilde{\pi}}_N\,\,,\,\, {\widetilde{\chi}}_{i+4} \equiv {\widetilde{\pi}}^i\,\,.
\label{third}
\end{eqnarray}

The inverse of the second class constraint matrix, ${\widetilde{C}}_{\alpha\beta}\equiv \{{\widetilde{\chi}}_{\alpha},{\widetilde{\chi}}_{\beta}\}$, is still 
\begin{equation}
{\widetilde{C}}^{-1}_{\alpha\beta}\equiv 
 \left(
     \begin{array}{c c c c}
      0 & 0 & -1 & 0 \\
    0 & 0 &  0 & -\mathbb{I} \\
    1 & 0 & 0  &  0   \\
    0 & \mathbb{I} & 0 & 0
      \end{array}
    \right)\,,
    \label{sectilde}
\end{equation}
as in the JF \eqref{matrix}.  
Now the new total Hamiltonian ${\widetilde{H}}_{T}^{'}$ is 
\begin{equation}
{\widetilde {H}}^{'}_{T}=\int d^{3}x \left({\widetilde{\lambda}}^{N} {\widetilde{\pi}}_N + {\widetilde{\lambda}}_{i}{\widetilde{\pi}}^{i}+{\widetilde{N}}{\widetilde{{\mathcal{H}}}}{}'+{\widetilde{N}}_i{\widetilde{{\mathcal{H}}}}^{i}{}' \right)\,. 
\label{hamiltonianatotEFp}
\end{equation}

Imposing that the gauge fixing constraints \eqref{third} be preserved on the constraint surface, we get 
\begin{equation}
    \dot{\widetilde{\chi}}_0(x)\approx\left\{\widetilde{N}(x)-c_0(16\pi G\phi(x))^{\frac{1}{2}},{\widetilde{H}}_T^{'}  \right\}\approx 0\,, 
    \label{condo}
\end{equation}
which implies 
\begin{equation}
{\widetilde{\lambda}}^{N}(x)\approx 0\, .
\label{lagrange}
\end{equation}

In a similar way imposing
\begin{equation}
\dot{\widetilde{\chi}}_{i}(x)\approx\left\{{\widetilde{N}}_i(x)-c_i(16\pi G\phi(x)),{\widetilde{H}}_T^{'}  \right\}\approx 0\,,
\label{condo1}
\end{equation}
it follows
\begin{equation}
{\widetilde{\lambda}}_{i}(x)\approx 0\, .
\label{lagrange1}
\end{equation}
The remaining second class constraints ${\widetilde{\chi}}_4 \equiv {\widetilde{\pi}}_N\,\,,\,\, {\widetilde{\chi}}_{i+4} \equiv {\widetilde{\pi}}^i$ are automatically preserved, as it is easy to see. 
Therefore the total Hamiltonian ${\widetilde{H}}_T^{'}$ reduces to the ADM Hamiltonian $\widetilde{H}^{'}_{ADM}\equiv \int d^{3}x 
\left({\widetilde{N}}\widetilde{{\mathcal{H}}}{}'+{\widetilde{N}}_i{\widetilde{{\mathcal{H}}}}^{i}{}' \right)$. 

We define the Dirac's brackets following \eqref{Diracbrackets1}, where, now, the inverse of second class constraint matrices ${\widetilde{C}}^{-1}_{\alpha\beta}$ is given by \eqref{sectilde}, and immediately, by the very definition of first class constraints, observe, as for \eqref{equivo}, 
\begin{eqnarray}
\{{\widetilde{\mathcal{H}}}^{'}_i(x) ,{\widetilde{\mathcal{H}}}^{'}_j(x')\}_{DB}&=& \{{\widetilde{\mathcal{H}}}^{'}_i(x) ,{\widetilde{\mathcal{H}}}^{'}_j(x')\}, \nonumber\\
\{{\widetilde{\mathcal{H}}}^{'}_i(x) ,{\widetilde{{\mathcal{H}}}}^{'}(x')\}_{DB}&=&\{{\widetilde{\mathcal{H}}}^{'}_i(x) ,{\widetilde{{\mathcal{H}}}}^{'}(x')\}\, , \nonumber\\
\{{\widetilde{\mathcal{H}}}^{'}(x) ,{\widetilde{\mathcal{H}}}^{'}(x')\}_{DB}&=&\{{\widetilde{\mathcal{H}}}^{'}(x) ,{\widetilde{\mathcal{H}}}^{'}(x')\}\, .
\label{equivo1}
\end{eqnarray}

Now we can write the equations of motion, in the Einstein frame, using Dirac's brackets
\begin{eqnarray}
\dot{\widetilde{h}}_{ij}&\approx&\{{\widetilde{h}}_{ij},\widetilde{H}^{'}_{ADM}\}_{DB}\nonumber\\
 &\approx&{\widetilde{D}}_i {\widetilde{N}}_j+{\widetilde{D}}_j{\widetilde{N}}_i+ \frac{(32\pi G) \widetilde{N}}{ \sqrt{\widetilde{h}}} \left({\widetilde{\pi}}_{ij}-\frac{{\widetilde{\pi}}_h}{2} \widetilde{h}_{ij} \right), \nonumber \\
\label{metricoc2}
\end{eqnarray}
and 
\begin{eqnarray}
\dot{\widetilde{\pi}}^{ij} &\approx& \{{\widetilde{\pi}}^{ij}, \widetilde{H}^{'}_{ADM}\}_{DB}\nonumber\\
&\approx& 
-\frac{\widetilde{N} \sqrt{\widetilde{h}}}{16\pi G}\left ({}^{(3)}{\widetilde{R}}^{ij}-\frac{1}{2}{}^{(3)}{\widetilde{R}}\,{\widetilde{h}}^{ij}\right) \nonumber \\
&&+\frac{(16\pi G) \widetilde{N}}{2\sqrt{\widetilde{h}}}\widetilde{h}^{ij}\left( {\widetilde \pi}^{pk}{\widetilde \pi}_{pk}-\frac{{{\widetilde \pi}_h}^2}{2}\right) \nonumber\\
&&-2\frac{(16\pi G) \widetilde{N}}{\sqrt{\widetilde{h}}}\left(\widetilde{\pi}^{ik}\widetilde{\pi}_k^{~j}-\frac{1}{2}\widetilde{\pi}_h \widetilde{\pi}^{ij}\right) -\widetilde{h}^{ij}\frac{\sqrt{{\widetilde h}}\widetilde{N}}{2} V(\phi)\nonumber \\
&&+\frac{\sqrt{\widetilde{h}}}{16\pi G}\left(\widetilde{D}^i\widetilde{D}^j \widetilde{N} - \widetilde{h}^{ij}\widetilde{D}^k\widetilde{D}_k \widetilde{N}\right)\nonumber\\
&&+\sqrt{\widetilde{h}} \widetilde{D}_k \left(\frac{1}{\sqrt{\widetilde{h}}} \widetilde{N}^k \widetilde{\pi}^{ij}\right) 
-\widetilde{\pi}^{ki}\widetilde{D}_k \widetilde{N}^j-\widetilde{\pi}^{kj}\widetilde{D}_k \widetilde{N}^i. \nonumber\\
\label{pidotequation}
\end{eqnarray} 
The equation of motion for $\phi$ is:
\begin{eqnarray}
\dot\phi &\approx& \{\phi, \widetilde{H}^{'}_{ADM}\}_{DB}\nonumber\\
 &\approx& \frac{8(\pi G) \phi^{2}}{\sqrt{\widetilde{h}}\left(\omega + \frac{3}{2}\right)}{\widetilde{\pi}_{\phi}}\widetilde{N}+\widetilde{N}_i {\widetilde D}^{i} \phi\,.
\label{phiequation1}
\end{eqnarray}
Finally, the evolution equation for ${\widetilde{\pi}}_{\phi}$ results to be
\begin{eqnarray}
{\dot{\widetilde{\pi}}}_\phi &\approx& \{{{\widetilde{\pi}}}_\phi, \widetilde{H}^{'}_{ADM}\}_{DB}\nonumber\\
 &\approx & 
 -\frac{\sqrt{\widetilde{h}}}{8\pi G}\left(\omega+\frac{3}{2}\right)\frac{\widetilde{N}}{\phi^3}\widetilde{D}^{i}\phi\widetilde{D}_{i}\phi  \nonumber\\
&&+\frac{\sqrt{\widetilde{h}}}{8\pi G}\frac{(\omega+\frac{3}{2})}{\phi^2}\widetilde{D}^{i}\widetilde{N}\widetilde{D}_{i}\phi \nonumber\\
&&+\frac{\sqrt{\widetilde{h}}}{8\pi G}\left(\omega+\frac{3}{2}\right)\frac{\widetilde{N}}{\phi^2}\widetilde{D}^{i}\widetilde{D}_{i}\phi  \nonumber\\
&&-\frac{8\widetilde{N}(\pi G)\phi{{\widetilde{\pi}}}^{2}_\phi}{{\sqrt{\widetilde{h}}\left(\omega+\frac{3}{2}\right)}}+
\widetilde{D}^i\left(\widetilde{N}_i{{\widetilde{\pi}}}_\phi\right) \nonumber\\
&&-\sqrt{\widetilde{h}} \widetilde{N} \frac{dV(\phi)}{d\phi}\,.
\label{piphifinale}
\end{eqnarray}

Strongly imposing the second class constraints 
defined in  \eqref{third},  
we eliminate ($\widetilde{N}\,,{\widetilde{N}}_{i}\,,{\widetilde{\pi}}_N\,,{\widetilde{\pi}}^i$) as dynamical variables. 
Substituting the constraints into the previous equations of motion \eqref{metricoc2}-\eqref{piphifinale}  we get:
\begin{eqnarray}
\dot{\widetilde{h}}_{ij}\approx
16 \pi G\left(c_j {\widetilde{D}}_i \phi
+c_i  {\widetilde{D}}_j \phi\right)\,,
 \label{hDB}    
\end{eqnarray}
\begin{eqnarray}
\dot{\widetilde{\pi}}^{ij} &\approx& 
-\frac{c_0\left(16\pi G \phi\right)^{1\over 2} \sqrt{\widetilde{h}}}{16\pi G}\left ({}^{(3)}{\widetilde{R}}^{ij}-\frac{1}{2}{}^{(3)}{\widetilde{R}}\,{\widetilde{h}}^{ij}\right) \nonumber \\
&&+\frac{(16\pi G) c_0\left(16\pi G \phi\right)^{1\over 2}}{2\sqrt{\widetilde{h}}}\widetilde{h}^{ij}\left( {\widetilde \pi}^{pk}{\widetilde \pi}_{pk}-\frac{{{\widetilde \pi}_h}^2}{2}\right) \nonumber\\
&&-2\frac{(16\pi G) c_0\left(16\pi G \phi\right)^{1\over 2}}{\sqrt{\widetilde{h}}}\left(\widetilde{\pi}^{ik}\widetilde{\pi}_k^{~j}-\frac{1}{2}\widetilde{\pi}_h \widetilde{\pi}^{ij}\right) \nonumber\\
&&-\widetilde{h}^{ij}\frac{\sqrt{{\widetilde h}} c_0\left(16\pi G \phi\right)^{1\over 2}}{2} V(\phi)\nonumber \\
&&+\frac{c_0\sqrt{\widetilde{h}}}{2\left(16\pi G\phi\right)^{1\over 2}}
\left[-\frac{\widetilde{D}^i \phi \widetilde{D}^j\phi}{2\phi}+\widetilde{D}^i\widetilde{D}^j \phi\right.\nonumber\\
&&\left. - \widetilde{h}^{ij}\left(-\frac{\widetilde{D}^k \phi \widetilde{D}_k\phi}{2\phi}+\widetilde{D}^k\widetilde{D}_k \phi\right)\right]\nonumber\\
&&+(16\pi G)\sqrt{\widetilde{h}} c^k \widetilde{D}_k \left(\frac{1}{\sqrt{\widetilde{h}}} \phi \widetilde{\pi}^{ij}\right)\nonumber\\ 
&&-(16\pi G)\widetilde{\pi}^{ki} c^j \widetilde{D}_k \phi
-(16\pi G)\widetilde{\pi}^{kj}c^i\widetilde{D}_k \phi\,, 
\label{piDB}
\end{eqnarray} 
\begin{eqnarray}
\dot\phi\approx \frac{c_0 (16\pi G \phi)^{3\over 2} }{2\sqrt{\widetilde{h}}\left(\omega + \frac{3}{2}\right)}{\widetilde{\pi}_{\phi}}\phi+\left( 16\pi G \phi\right)c_i {\widetilde D}^{i} \phi\,,
\label{phiDB}
\end{eqnarray}
\begin{eqnarray}
{\dot{\widetilde{\pi}}}_\phi 
 &\approx & 
 -\frac{c_0\sqrt{\widetilde{h}}}{(16\pi G\phi)^{1\over2}\phi^2}\left(\omega+\frac{3}{2}\right)
 \widetilde{D}^{i}\phi\widetilde{D}_{i}\phi  \nonumber\\
&&+\frac{2c_0\sqrt{\widetilde{h}}}{(16\pi G\phi)^{1\over2}\phi}\left(\omega+\frac{3}{2}\right)\widetilde{D}^{i}\widetilde{D}_{i}\phi  \nonumber\\
&&-\frac{c_0 (16 \pi G \phi)^{3\over 2} {{\widetilde{\pi}}}^{2}_\phi}{{2\sqrt{\widetilde{h}}\left(\omega+\frac{3}{2}\right)}}
+16\pi G c_i\widetilde{D}^i\left(\phi{{\widetilde{\pi}}}_\phi\right) \nonumber\\
&&-c_0 \sqrt{\widetilde{h}} \left(16\pi G \phi\right)^{1\over 2} \frac{dV(\phi)}{d\phi}\,.
\label{piphifinaleDB}
\end{eqnarray}

Similarly to the flat FLRW case, the lapse and the shifts - and their conjugate momenta - are not anymore dynamical variables. 
Therefore, the Hamiltonian transformation from the JF to the EF frames \eqref{JFEFtrans2a}-\eqref{JFEFtrans2b}
is reduced to fewer dynamical variables, on which it is completely canonical 
(see Eqs. \eqref{coniugate}-\eqref{noncanonicalcond2}).

\section{Discussion and Conclusions}
\label{Conclusions}

The aim of this article is to clarify several issues and assumptions present in the literature.

Firstly, we stressed that it is not true that the Hamiltonian transformations from the Jordan to the Einstein frame are canonical on the extended phase-space. 
In order to make them Hamiltonian canonical,
we have to gauge-fix the lapse function $N$ and the shifts functions $N_i$. 
This gauge-fixing makes the primary first-class constrains second class and they can be solved provided that the Dirac's brackets are defined. 
On the reduced phase-space obtained in this way the Hamiltonian transformations from the Jordan to the Einstein frame
are canonical.

If two classical theories are related by a canonical transformation, the symplectic two form is preserved. This is equivalent to say that the Poisson brackets among phase-space variables remains unchanged. 
Two Hamiltonian theories connected by canonical transformations are not necessarily physical equivalent. We show two examples of Hamiltonian canonical transformation between two physically different systems.
 
In fact, if we consider the case of an {\em uni-dimensional harmonic oscillator}
of mass $m$ and frequency $\omega$ the Hamiltonian function is:
\begin{equation}
H=\frac{p^2}{2m} + \frac{m\omega^2}{2} q^2\,.
\label{classicalhami}
\end{equation}
If we apply the following Hamiltonian canonical transformation \cite{goldstein2002classical,sussman2001structure}
\begin{eqnarray}
\label{canoclas1}
q&=&\sqrt{\frac{2P}{m\omega}}\sin Q\,,\\
p&=&\sqrt{2m\omega P} \cos Q \,,
\label{canoclas2}
\end{eqnarray}
the Hamiltonian function \eqref{classicalhami} in the new variables $(Q,P)$ is:
\begin{equation}
 H=\omega P \, .
 \label{classhamiN}
\end{equation}
Since the energy, in this system, is conserved, and then a constant of motion $E$, we get: 
\begin{equation}
P=\frac{E}{\omega}\, . 
\label{constanto}
\end{equation}
Therefore
\begin{equation}
\dot{Q}=\frac{\partial H}{\partial P}=\omega \, ,
\label{simple}
\end{equation}
then 
\begin{equation}
Q=\omega t + \alpha \, ,  
\label{constantvelo}
\end{equation}
$\alpha$ being an integration constant. Replacing in  \eqref{canoclas1} we obtain:
\begin{equation}
q(t)=\sqrt{\frac{2E}{m\omega^2}} \sin(\omega t + \alpha)\,.
\end{equation}

It is well know that symmetries of a mechanical system are the generators of Hamiltonian canonical transformations \cite{goldstein2002classical}. They turn to be automorphisms, which  preserve the structure of the equations of motion. 
The transformations \eqref{canoclas1}-\eqref{canoclas2}  are not a symmetries of the system, so they do not map a physical system into an equivalent one, from a physical point of view. 
In fact \eqref{canoclas1}-\eqref{canoclas2} map the harmonic oscillator \eqref{classicalhami} 
into a particle moving with constant velocity \eqref{constantvelo}. 
Transformations \eqref{canoclas1}-\eqref{canoclas2} change the physical system, mapping it into one whose equations of motion, cfr. Eq. \eqref{simple}, are easier to solve. 
This is also, in our opinion, what is happening, at the end, in the Hamiltonian transformations from the Jordan to the Einstein frame: we map our theory into the Einstein frame, where things get easier to solve. 

A second example is the Hamiltonian canonical transformations between the Jordan frame and the {\em anti-Newtonian} frame \cite{Galaverni:2021xhd}. 
This anti-Newtonian frame is defined by the anti-Newtonian (or {\em anti-gravity}) transformations \cite{Niedermaier2019,Niedermaier2020,Zhang2011,Zhou2012,Cruz:2018}.
In fact, \cite{Galaverni:2021xhd}
 the following set of anti-Newtonian (or anti-gravity) transformations 
\begin{eqnarray}
&&{{\widetilde {N}}}^{*}=N \, ,\, \, {\widetilde{\pi}}_{N^{*}}=\pi_N  \,,\nonumber\\
&&{{\widetilde {N^{*}_i}}}=N_i \, ,\, \, {\widetilde{{\pi}}}^{*}{}^{i}=\pi^i  \,,\nonumber\\
&&\, \widetilde{h}^{*}_{ij}=(16\pi G\phi)h_{ij} \, ,  \,\, 
{\widetilde{\pi}}^{*}{}^{ij}=\frac{\pi^{ij}}{(16\pi G\phi)^{1\over 2}}  \,,\nonumber\\
&&{{\widetilde {\phi}}}^{*}=\phi \, , \,\, {\widetilde \pi}^{*}_\phi=\frac{1}{\phi} ( \phi \pi_{\phi}-\pi_{h})\,,
\label{JFEFtrans3}
\end{eqnarray}
are Hamiltonian canonical transformations on the extended phase space without making any gauge-fixing. 

The physics of these transformations is synthesized, in two dimensions, by the following metric 
\begin{equation}
ds^{2}=-dt^{2}+\lambda^{2} dx^{2}\,.
\label{carrolian}
\end{equation}
When $\lambda >1$ this metric corresponds to a space-time where the limiting velocity is less than the velocity of light. The light-cone structure squeezes as $\lambda \gg 1$; which corresponds to a situation in which space-like distances enhance over time-like distances. 
In the limit $\lambda \to \infty$, we have that the limit velocity goes to zero ($c \to 0$). 
This corresponds to {\it{Carroll gravity}} and represents the case of strong gravitational fields, in which the gravitational constant $G$ becomes very large, $G \to \infty$, and the limit velocity vanishes, $c \to 0$ \cite{Niedermaier2019,Niedermaier2020}. Clearly, the Brans-Dicke theory in the Jordan frame and the corresponding theory in the {\it{anti-Newtonian}} frame represent two different system.

The gauge-fixed Hamiltonian transformations from the Jordan to the Einstein frames are canonical but only map solutions of the equations of motion in the Jordan frame into solutions of the equations of motion in the Einstein frame. 
Pairwise, the Hamiltonian canonical ``anti-Newtonian'' transformations map the solutions of the equations of motion of the Branse-Dicke theory in the Jordan frame into the solutions of the equations of motion of an alternative theory of gravity in the anti-gravity frames. 
The Jordan-Einstein frames transformations and the anti-Newtonian transformations can be seen as generators of solutions of the equations of motion. In fact,  a solution of the equations of motion in a frame could be used to derive a solution of the equations of motion of the correspondent theory in the related frame. 

These considerations suggest us to stand with part of the scientific community that are in favor of considering the Jordan frame as the physical frame. 
The question is still debated and a solution has not been found yet 
\cite{Faraoni:1999hp,Kamenshchik:2016gcy,Ruf2017,Nandi1,Nandi2,Nandi3,Pandlejee2016,Pandey:2016jmv,Barreto:2017lid,Paliathanasis:2023gfq}. 
At this point, the check of whether or not the physical observables, calculated separately in the Einstein and Jordan frame, reproduce the same result in both frames, should throw light on the physical equivalence of Jordan and Einstein frames. 
This very point is still quite controversial \cite{Deruelle2010h,Chiba:2013mha,Quiros:2012rnn,Capozziello2010,Bahamonde:2016wmz,Bahamonde:2017kbs,Francfort2019,Racioppi:2021jai,Ghosh:2022ppn,Rondeau:2017xck}. 
We have not tackled this topic in our analysis and we plan to discuss it in a future work.

\begin{acknowledgement}
We thank Jack Wisdom for useful discussions.
\end{acknowledgement}

\appendix

\section{Brans-Dicke theory in flat FLRW universe for $\omega = -3/2$}
\label{AppA1}

\subsection{Gauge fixing in the JF}
\label{AppA1a}

In the particular case $\omega=-3/2$ Brans--Dicke theory \eqref{BDaction} is invariant under Weyl conformal transformations.
As a consequence of this symmetry there is an additional primary constraint \cite{Galaverni:2021jcy,Galaverni:2021xhd}:
\begin{equation}
C_\phi\equiv \frac{1}{2}a\pi_{a}- \phi \pi_{\phi}\,.    
\end{equation}
The total Hamiltonian in the JF is:
\begin{eqnarray}
\label{H:JF:eq}
H_{\mathrm{T}}^{(-3/2)}&=&N\left[ - \frac{\pi_a^2}{24 a \phi}
+ a^3 U(\phi)\right]  + \lambda_N \pi_N+ \lambda_\phi C_\phi\nonumber\\
&=& N H^{(-3/2)} + \lambda_N \pi_N+ \lambda_\phi C_\phi\,.
\end{eqnarray}

After gauge fixing \eqref{JFgaugef}, we note that $C_\phi$ remains a first class constraint, since: 
$\{C_\phi,N-c_0\}\approx0$,
$\{C_\phi,H_{\mathrm{T}}^{(-3/2)}\}\approx\frac{N}{2}H^{(-3/2)}\approx0$ and 
$\{C_\phi,\pi_N\}\approx0$,
and similarly the Hamiltonian constraint: 
$\{H_{\mathrm{T}}^{(-3/2)},N-c_0\}\approx0$ and $\{H_{\mathrm{T}}^{(-3/2)},N-c_0\}\approx\pi_N$.

In this $\omega=-3/2$ case, we have the same two secondary constraints 
$\chi_0\equiv N-c_0\,,\;\mbox{and}\;\chi_1\equiv \pi_N$, see Eq.~\eqref{FLRW_JW_constr},
and the corresponding Dirac brackets \eqref{Diracbrackets1}.

We verify that, also considering the total Hamiltonian defined in Eq.~\eqref{H:JF:eq},

the systems remains on the reduces phase space defined by the secondary constraints \eqref{FLRW_JW_constr} if:
\begin{equation}
\dot{\chi}_1 \approx \left\{N-c_0, H_{\mathrm{T}}^{(-3/2)}\right\}\approx0\,,    
\end{equation}
which implies $\lambda_N\approx0$, 
and if:
\begin{equation}
\dot{\chi}_2\approx\left\{\pi_N, H_{\mathrm{T}}^{(-3/2)}\right\}\approx 0\,,    
\end{equation}
which is automatically verified.
$N$ and $\pi_N$ are not anymore dynamical variables, therefore, we have to evaluate only four equations of motion using the Dirac brackets:
\begin{eqnarray}
\label{adotJFeq}
\dot{a}&\approx&\left\{a,H_{\mathrm{T}}^{(-3/2)}\right\}_{DB}
%\nonumber\\       &=& 
\approx -\frac{N\pi_a}{12 a\phi}+\frac{\lambda_\phi a}{2}\;,
\end{eqnarray}
\begin{eqnarray}
\dot{\pi}_a &\approx&\{{\pi}_a,H_{\mathrm{T}}^{(-3/2)}\}_{DB}\nonumber\\
       &\approx&-\frac{N \pi_a^2}{24 \phi a^2}-3 N a^2 U(\phi)-\frac{\lambda_\phi\pi_a}{2}\,,
\end{eqnarray}
\begin{eqnarray}       
\dot{\phi}&\approx& \{\phi, H_{\mathrm{T}}^{(-3/2)}\}_{DB}\approx -\lambda_\phi\phi\,, 
\end{eqnarray}
\begin{eqnarray}
\dot{\pi}_\phi&\approx& \{\pi_{\phi},H_{\mathrm{T}}^{(-3/2)}\}_{DB}\nonumber\\
&\approx&
-N\frac{\pi_a^2}{24 a\phi^2}- N a^3  \frac{dU(\phi)}{d\phi}+\lambda_\phi \pi_\phi\nonumber\\
&\approx&
-N\frac{\pi_a^2}{24 a\phi^2}-\frac{2 N a^3 U(\phi)}{\phi}+\lambda_\phi \pi_\phi\,,
\label{piphidotJFeq}
\end{eqnarray}
where in the last line we used:
\begin{equation}
\label{eq:phi:eq}
\phi \frac{d U(\phi)}{d\phi} = 2 U(\phi)\,,    
\end{equation}
see Eq.~\eqref{eqstophi} for $\omega=-3/2$.

Strongly imposing the second class constraints $N=c_0$ and $\pi_N=0$, see Eqs. \eqref{FLRW_JW_constr}, we get the equations of motion on the reduces phase space in the particular case $\omega=-3/2$:
\begin{eqnarray}
\label{adotJFeq2}
 \dot{a}&\approx& -\frac{c_0\pi_a}{12 a\phi}+\frac{\lambda_\phi a}{2}\;, \\
 \dot{\pi}_a &\approx&-\frac{c_0 \pi_a^2}{24 \phi a^2}-3 c_0 a^2 U(\phi)-\frac{\lambda_\phi\pi_a}{2}\,,\\
 \dot{\phi}&\approx&-\lambda_\phi\phi\,, \\
 \dot{\pi}_\phi&\approx&-c_0\frac{\pi_a^2}{24 a\phi^2}-\frac{2 c_0 a^3 U(\phi)}{\phi}+\lambda_\phi \pi_\phi\,.
 \label{piphidotJFeq2}
\end{eqnarray}

\subsection{Gauge fixing in the EF}
\label{AppA1b}

Using the Weyl (conformal) transformations defined in \eqref{JFEFtrans}
it easy to pass from the total Hamiltonian in the JF, see Eq.~\eqref{H:JF:eq}, to the total Hamiltonian in the EF:
\begin{eqnarray}
\label{H:EF:eq1}
\widetilde{H}_T^{(-3/2)}&=&  \widetilde{N} \widetilde{a}^3  \left[
-\frac{2 \pi G \widetilde{\pi}_a^2}{3\widetilde{a}^4}
+V(\widetilde{\phi})
\right] + \widetilde{\lambda}_N \widetilde{\pi}_N +\lambda_\phi\widetilde{C}_\phi\nonumber\\
&=&\widetilde{N} \widetilde{H}^{(-3/2)} +\widetilde{\lambda}_N \widetilde{\pi}_N +\widetilde{\lambda}_\phi\widetilde{C}_\phi\,.
\end{eqnarray}
The additional primary constraint in the EF becomes:
\begin{equation}
 \widetilde{C}_\phi=-\phi \widetilde{\pi}_\phi\,.   
\end{equation}
Also in this case the gauge fixing in the EF is implemented introducing the secondary constraint:
\begin{equation}
\widetilde{N}-c_0 (16\pi G\phi)^{1\over 2}  \approx 0\,,    
\end{equation}
see also Eq.~\eqref{EFconstraint}.

Here, the Hamiltonian constraint remains first class, since:
\begin{equation}
\{\widetilde{H}^{(-3/2)}, \widetilde{N}-c_0 (16\pi G\phi)^{1\over 2} \}=0\,,    
\end{equation}
and also:
\begin{equation}
\label{HCEF}
\{\widetilde{H}^{(-3/2)}, \widetilde{C}_\phi \}=- \widetilde{a}^3 \phi \frac{dV(\phi)}{d\phi}=0\,.    
\end{equation}
since in the EF $\phi\frac{dV(\phi)}{d\phi}=0$.

On the contrary, $\widetilde{C}_\phi$ appears now to be second class, due to the gauge fixing constraint:
\begin{equation}
\{\widetilde{N}-c_0 (16\pi G\phi)^{1\over 2},  \widetilde{C}_\phi \}=
\frac{c_0 (16\pi G\phi)^{1\over 2}}{2}\,.
\end{equation}
However, it is always possible to redefine the conformal constraint as:
\begin{equation}
 \widetilde{C}_\phi-\frac{c_0 (16\pi G\phi)^{1\over 2}}{2}\widetilde{\pi}_N\,,    
\end{equation}
obtaining a first class constraint:
\begin{equation}
\{\widetilde{N}-c_0 (16\pi G\phi)^{1\over 2}, \widetilde{C}_\phi-\frac{c_0 (16\pi G\phi)^{1\over 2}}{2}\widetilde{\pi}_N  \}\approx 0\,.   
\end{equation}

Therefore, the total Hamiltonian introduced in Eq.~\eqref{H:EF:eq1}, in this case, is now redefined as:
\begin{eqnarray}
\label{H:EF:eq2}
\widetilde{H}_T^{\prime(-3/2)}&=&\widetilde{N} \widetilde{H}^{(-3/2)} +\widetilde{\lambda}_N \widetilde{\pi}_N \nonumber\\
&&+\widetilde{\lambda}_\phi\left[ -\phi \widetilde{\pi}_\phi -\frac{c_0 (16\pi G\phi)^{1\over 2}}{2}\widetilde{\pi}_N    \right]\, .
\end{eqnarray}

Also in this case, the only two irreducible second class constraints in EF are:
$\widetilde{\chi}_0\equiv \widetilde{N}-c_0(16\pi G\phi)^{1\over 2}$ and 
$\widetilde{\chi}_1\equiv \widetilde{\pi}_N$, see Eq.~\eqref{FLRW_EF_constr}.
The  definition of the Dirac brackets coincides with the one of the $\omega\neq-3/2$ case, see Sec.~\ref{FLRW:EF:neq}.
Evolution remains confined in the reduced phase space defined secondary constraints if:
\begin{equation}
\dot{\widetilde{\chi}}_0\approx\left\{\widetilde{N}-c_0(16\pi G\phi)^{1\over 2},{\widetilde{H}}_T^{\prime(-3/2)}  \right\}\approx0\,,  
\end{equation}
which implies $ \widetilde{\lambda}_N\approx0$, and if:
\begin{equation}
\dot{\widetilde{\chi}}_1\approx\left\{\widetilde{\pi}_N,{\widetilde{H}}_T^{\prime(-3/2)}  \right\}\approx0\,,    
\end{equation}
which is automatically verified.

We have to consider only four equations of motion evaluated using the Dirac brackets:
\begin{eqnarray}
\label{adotEFeq}
\dot{\widetilde{a}}&\approx&\left\{\widetilde{a},\widetilde{H}_T^{\prime(-3/2)}\right\}_{DB}\approx -\widetilde{N} \frac{4\pi G \widetilde{\pi}_a }{3 \widetilde{a}}\,, 
\end{eqnarray}
\begin{eqnarray}
\dot{\widetilde{\pi}}_a&\approx&\left\{\widetilde{\pi}_a,\widetilde{H}_T^{\prime(-3/2)}\right\}_{DB}\nonumber\\
    &\approx& \widetilde{N} 
\left[-\frac{(2\pi G) \widetilde{\pi}_a^2 }{3 \widetilde{a}^2}
-3 \widetilde{a}^2 V(\phi) \right]
\,,
\end{eqnarray} 
\begin{eqnarray}
\dot{\widetilde{\phi}}&\approx&\left\{\widetilde{\phi},\widetilde{H}_T^{\prime(-3/2)}\right\}_{DB}\approx -\widetilde{\lambda}_\phi\widetilde{\phi}\,,
\end{eqnarray}
\begin{eqnarray}
\dot{\widetilde{\pi}}_\phi&\approx&\left\{\widetilde{\pi}_\phi,\widetilde{H}_T^{\prime(-3/2)}\right\}_{DB}\nonumber\\
&\approx&\left\{\widetilde{\pi}_\phi,\widetilde{H}_T^{\prime(-3/2)}\right\}\nonumber\\
&&- \left\{\widetilde{\pi}_\phi, \widetilde{N}-c_0(16\pi G\phi)^{1\over 2}\right\} C^{-1}_{01}\left\{\widetilde{\pi}_N,\widetilde{H}_T^{\prime(-3/2)}  \right\}\nonumber\\
    &\approx& -\widetilde{N} \widetilde{a}^3 \frac{d V(\phi)}{d\phi}
+\widetilde{\lambda}_\phi \widetilde{\pi}_\phi-\frac{\widetilde{\lambda}_\phi c_0}{2} \frac{16 \pi G}{(16\pi G\phi)^{1\over 2}}\widetilde{\pi}_N\nonumber\\
   &&+ \frac{c_0}{2} \frac{16 \pi G}{(16\pi G\phi)^{1\over 2}} \widetilde{H}^{(-3/2)}
      \,.
\label{piphidotEFeq}
\end{eqnarray}

Strongly imposing the second class constraints $\widetilde{N}= c_0  (16\pi G\phi)^{1\over 2}  $ and $\widetilde{\pi}_N=0$, see Eq.~\eqref{FLRW_EF_constr},  we get the equations of motion on the reduced phase space:
\begin{eqnarray}
\label{adotEFeq2}
\dot{\widetilde{a}}&\approx& -c_0(16\pi G\phi)^{1\over 2} \frac{4\pi G \widetilde{\pi}_a }{3 \widetilde{a}}\,,\\  
\dot{\widetilde{\pi}}_a&\approx& c_0(16\pi G\phi)^{1\over 2} 
\left[-\frac{(2\pi G) \widetilde{\pi}_a^2 }{3 \widetilde{a}^2}
-3 \widetilde{a}^2 V(\phi) \right],\,\,\,\\
\dot{\widetilde{\phi}}&\approx&-\widetilde{\lambda}_\phi\widetilde{\phi}\,,\\
\dot{\widetilde{\pi}}_\phi &\approx& \widetilde{\lambda}_\phi \widetilde{\pi}_\phi\,.
\label{piphidotEFeq2}
\end{eqnarray}

In this $\omega=-3/2$ case too, the equations of motion in the EF \eqref{adotEFeq2}-\eqref{piphidotEFeq2}
can be transformed in the equations in the JF \eqref{adotJFeq2}-\eqref{piphidotJFeq2} -- and vice versa --
using the Weyl (conformal) transformation  \eqref{JFEFtrans}.

This is a clear consequence of the Hamiltonian canonical equivalence between JF and EF on the reduced phase space.

\section{Brans-Dicke theory in ADM spacetime for $\omega=-3/2$}
\label{App2}

\subsection{Gauge fixing in the JF}
\label{App2a}
In the ADM field theory case too, there is and additional first class constraint $C_{\phi}$ due to invariance under Weyl (conformal) transformations
for $\omega=-3/2$. 
The total Hamiltonian in the JF is \cite{Esposito1992,Galaverni:2021xhd}:
\begin{eqnarray}
H_{T}^{(-3/2)}&=&\int d^{3}x \left(\lambda^N \pi_N + \lambda_{i}\pi^{i} + \lambda_{\phi}C_{\phi}\right.\nonumber\\
&&\left.+N{\mathcal{H}^{(-3/2)}}+N_i{{\mathcal{H}}^{i}}^{(-3/2)} \right)\;\,, 
\end{eqnarray}
where $\lambda^N=\lambda^N(x)$, $\lambda^{i}(x)$, and $\lambda_{\phi}(x)$ are Lagrange multipliers, 
$\pi_N$, $\pi^{i}$ and $C_\phi$ are primary Dirac's constraints, $C_\phi$ being:
\begin{equation}
C_\phi=\pi^{ij}h_{ij}-\phi \pi_\phi\,.
\label{conformo1}
\end{equation}
The Hamiltonian constraint $\mathcal{H}$, see \cite{Galaverni:2021xhd}, is 
\begin{eqnarray}
{\mathcal H}^{(-3/2)}&=&{\sqrt{h}}\Bigg\{ \left[-\phi\;  {}^{3}R+\frac{1}{\phi h}\left( \pi^{ij}\pi_{ij}-\frac{{\pi_h}^2}{2}\right)\right] \nonumber\\
&&- \frac{3}{2\phi}D_i\phi D^i\phi+2D^iD_i\phi+U(\phi) \Bigg\}\,,
\end{eqnarray}
and the momentum constraints ${{\mathcal {H}}^i}^{(-3/2)}$
\begin{equation}
{{\mathcal {H}}^i}^{(-3/2)}= -2D_j\pi^{ji}+D^i\phi \pi_{\phi}\,. 
\end{equation}

As we stressed several times the gauge-fixing conditions, also for the $\omega=-\frac{3}{2}$ Brans-Dicke theory,
in the JF are
\begin{equation}
N=c_0\;\;\;\; N_{i}=c_i 
\label{JFimpongo1}\,\,\,,
\end{equation}
 which are implemented as secondary constraints:
\begin{equation}
\chi_0\equiv N-c_0 \approx 0\;\;,\;\chi_i\equiv N_i -c_i \approx 0\;.
\label{secondario1}
\end{equation}
The previously correspondent primary first class constraints \cite{Gionti2021} 
\begin{equation}
\chi_4\equiv\pi_{N}\approx 0 \;\;,\; \chi_{i+
4}\equiv\pi^{i}\approx 0 \;,
\label{primary-first1}
\end{equation}
become second class constraints since: 
\begin{eqnarray}
    \{ N(x)-c_0,\pi_N(x')\}\approx\delta^{(3)}(x-x') \;\;\nonumber\\ 
    \{N_i (x) - c_i, \pi^{j}(x')\}=\delta^{j}_{i}\delta^{(3)}(x-x')\,.
    \label{secondary1}
\end{eqnarray}

Imposing the second class constraints to be preserved we have:
\begin{equation}
 \dot{\chi}_0\approx \{N-c_0,H_T \}\approx 0\,,   
\end{equation}
which implies $\lambda^N(x)\approx0$, and: 
\begin{equation}
 \dot{\chi}_i\approx \{N_i-c_i,H_T \}\approx 0\,,   
\end{equation}
which implies $\lambda_i(x)\approx0$.
The other two second class constraints:
\begin{eqnarray}
 \dot{\chi}_4&\approx& \{\pi_N,H_T \}\approx 0\,,\\
 \dot{\chi}_i+4&\approx& \{\pi^i,H_T \}\approx 0\,, 
\end{eqnarray}
are automatically preserved.

The equations of motion, calculated with the Dirac's Brackets and substituting the second class constraints \eqref{secondario1}-\eqref{primary-first1} imposed strongly, are: 
\begin{eqnarray}
\dot{h}_{ij}&\approx& \{ h_{ij}, H_T \}_{DB}\nonumber\\ 
&\approx& \lambda_{\phi}h_{ij}+\frac{2c_0}{\phi \sqrt{h}} \left(\pi_{ij}-\frac{\pi_h}{2} h_{ij} \right)\,,
\label{metrico3}
\end{eqnarray}
\begin{eqnarray}
{\dot{\pi}}^{ij}&\approx&\{\pi^{ij}, H_T\}_{DB}\nonumber\\
&\approx& -\lambda_{\phi}\pi^{ij}-c_0 \sqrt{h}\phi \left( {}^{(3)}R^{ij}- \frac{h^{ij}}{2}\, {}^{(3)}R \right) \nonumber\\
&&+c_0 \sqrt{h}\left( D^i D^j-h^{ij} D^k D_k\right)(\phi) \nonumber \\
&&+\frac{c_0 h^{ij}}{2 \phi\sqrt{h}}\left( \pi^{ij}\pi_{ij}-\frac{{\pi}^{2}_{h}}{2}\right)+\frac{c_0}{\phi\sqrt{h}}\pi_h \pi^{ij}\nonumber\\
&&-\sqrt{h}\frac{c_0 \omega}{\phi}D^i \phi D^j\phi-\sqrt{h}\frac{c_0 \omega}{2\phi}h^{ij}D_k\phi D^k\phi \nonumber \\
&& -\frac{\sqrt{h}}{2} h^{ij} c_0 U(\phi) \nonumber\\
&&+c^k D_k \left(\pi^{ij} \right)-\frac{2c_0}{\sqrt{h}\phi}\pi^{iq}\pi_{q}^{\,\,j} \,,
\label{pilunga3}
\end{eqnarray}
\begin{eqnarray}
\dot \phi &&\approx \{\phi, H_T\}_{DB}\approx -\lambda_{\phi}\phi + c^i D_i \phi\,,
\label{phiequation3}
\end{eqnarray}
\begin{eqnarray}
{\dot{\pi}}_{\phi}&\approx&  \{{\pi}_{\phi}, H_T\}_{DB}\nonumber\\
&\approx& \lambda_{\phi}\pi_{\phi}+ c_0\sqrt{h}\, {}^{(3)}R +\frac{c_0}{{\phi}^2 \sqrt{h}}
\left(\pi^{ij}\pi_{ij}-\frac{{\pi_h}^2}{2} \right) \nonumber \\
&&+c_0\frac{\sqrt{h}\omega}{\phi^2}D_i\phi D^i\phi + 2c_0\sqrt{h}D_i \left(\frac{\omega}{\phi} D^{i}\phi\right) \nonumber\\
&&-c_0\sqrt{h}\frac{dU}{d\phi}+c^{i}D_i(\pi_{\phi})\,.
\label{equopiphi3}
\end{eqnarray}

\subsection{Gauge fixing in the EF}
\label{App2b}

Under Weyl (conformal) transformations, we easily obtain the total Haimltonian in the EF \cite{Galaverni:2021xhd}:
\begin{eqnarray}
{\widetilde H}_T^{(-3/2)}&=&\int d^{3}x\Big({\widetilde{\lambda}}^N{\widetilde \pi}_N+{\widetilde \lambda}^{i} {\widetilde \pi} _{i}+{\widetilde{\lambda}}_{\phi}\widetilde{C}_{\phi}\nonumber\\
&&+
{\widetilde{N}}{\widetilde{{\mathcal {H}}}}^{(-3/2)}+{\widetilde{N}}^{i}{\widetilde{{\mathcal {H}}}}_{i}^{(-3/2)}\Big),
\end{eqnarray}
here we have, see always \cite{Galaverni:2021xhd}:
\begin{equation}
{\widetilde{C}}_{\phi}=-\phi \widetilde{\pi}_{\phi}\,,
\label{conformalend}
\end{equation}
\begin{eqnarray}
{\widetilde{\mathcal H}}^{(-3/2)}&=&\frac{\sqrt{{\widetilde h}}}{16\pi G}\left[ -{}^{3}{\widetilde R}+\frac{(16\pi G)^2}{\widetilde h}\big( {\widetilde \pi}^{ij}{\widetilde \pi}_{ij}-\frac{{{\widetilde \pi}_h}^2}{2}\big)\right]\nonumber\\
&&+\sqrt{{\widetilde h}} V(\phi)\,, 
\label{hamiltoEF}
\end{eqnarray}
and:
\begin{equation}
 \widetilde{{\mathcal H}}_{i}^{(-3/2)}=-2{\widetilde D}_j{\widetilde \pi}^{j}_{i} \,.
\label{momentaEF}
\end{equation}

As remarked in the case $\omega \neq -\frac{3}{2}$, see Eq.~\eqref{EFimpongo1}, the gauge-fixing conditions happen to be:
\begin{equation}
\widetilde{N}=c_0\left( 16\pi G \phi\right)^{1\over 2}\;\;,\;\; {\widetilde{N}}_{i}=c_i\left( 16\pi G \phi\right)\,,
\label{EFimpongo2}
\end{equation}
and, as usual, we implement them as secondary Dirac's constraints: 
\begin{equation}
\widetilde{N}-c_0\left( 16\pi G \phi\right)^{1\over 2} \approx 0\;\;,\;\; {\widetilde{N}}_{i}-c_i\left( 16\pi G \phi\right)\, \approx 0\,.
\label{EFimpongo3}
\end{equation}

They continue to make, as in the previous cases (see Eq.~\eqref{secondary} and Eq.~\eqref{secondaryEF}), the primary first-class constraints, $\widetilde{\pi}_N (x) \approx 0$ and $\widetilde{\pi}^{j}(x) \approx 0$, second class: 
\begin{eqnarray}
    &&\{\widetilde{N}(x)-c_0\left( 16\pi G \phi\right)^{1\over 2},\widetilde{\pi}_N (x')\}\approx \delta^{(3)}(x-x') \;,\; \nonumber\\
     &&\{{\widetilde{N}}_i (x) - c_i\left( 16\pi G \phi (x) \right), \widetilde{\pi}^{j}(x')\}\approx \delta^{j}_{i} \delta^{(3)}(x-x').\nonumber\\
    \label{secondaryEF1}
\end{eqnarray}

It is quite straightforward to check that, contrary to what happens in the case $\omega\neq -\frac{3}{2}$ in the Einstein frame, see Sec.~\ref{EF1}, Hamiltonian and momentum constraints are clearly first class (there is no need of a re-definition):
\begin{eqnarray}
    &&\{\widetilde{N}(x)-c_0\left( 16\pi G \phi\right)^{1\over 2},{\widetilde{\mathcal{H}}}^{(-3/2)} (x')\}\approx 0 \;,\; \nonumber\\
     &&\{{\widetilde{N}}_i (x) - c_i\left( 16\pi G \phi (x) \right), {\widetilde{\mathcal{H}}}^{(-3/2)} (x')\}\approx 0 \nonumber\\
      &&\{\widetilde{N}(x)-c_0\left( 16\pi G \phi\right)^{1\over 2},{\widetilde{\mathcal{H}}}_i^{(-3/2)}(x')\}\approx 0 \;,\; \nonumber\\
     &&\{{\widetilde{N}}_i (x) - c_i\left( 16\pi G \phi (x) \right), {\widetilde{\mathcal{H}}}_i^{(-3/2)}(x')\}\approx 0 \,.  
    \label{secondaryEF2}
\end{eqnarray}

But, we notice the following:
\begin{eqnarray}
    &&\{\widetilde{N}(x)-c_0\left( 16\pi G \phi\right)^{1\over 2},{\widetilde{C}}_{\phi}\}\approx \frac{c_0\left( 16\pi G \phi\right)^{1\over 2}}{2} \;,\; \nonumber\\
    &&\{{\widetilde{N}}_i (x) - c_i\left( 16\pi G \phi (x) \right),{\widetilde{C}}_{\phi}\}\approx c_i\left( 16 \pi G \phi\right) ,
    \label{secondclassfin}
\end{eqnarray}
that is ${\widetilde{C}}_{\phi}$ looks, ``apparently'', second-class Dirac's constraint. 
If we re-define ${\widetilde{C}}_{\phi}$ as 
\begin{equation}
{\widetilde{C}}'_{\phi}\equiv -\phi \widetilde{\pi}_{\phi} -\frac{c_0\left( 16\pi G \phi\right)^{1\over 2}}{2} \widetilde{\pi}_N - c_i\left( 16 \pi G \phi\right)\widetilde{\pi}^{i}, 
\label{Cphiestendo}
\end{equation}
${\widetilde{C}}'_{\phi}$ stays first class as it is quite easy to check. 

Therefore, the new total Hamiltonian ${{\widetilde H}}_T^{'(-3/2)}$ is: 
\begin{eqnarray}
{\widetilde H}_T^{'(-3/2)}&=&\int d^{3}x\Big({\widetilde{\lambda}}^N{\widetilde \pi}_N+{\widetilde \lambda}^{i} {\widetilde \pi} _{i}+{\widetilde{\lambda}}_{\phi}\widetilde{C}'_{\phi}\nonumber\\
&&+
{\widetilde{N}}{\widetilde{{\mathcal {H}}}}^{(-3/2)}+{\widetilde{N}}^{i}{\widetilde{{\mathcal {H}}}}_{i}^{(-3/2)}\Big).
\end{eqnarray}

As we did in the Sec.~\ref{EF1}, we re-define the second class constraints as:
\begin{eqnarray}
&&{\widetilde{\chi}}_0 \equiv\widetilde{N}-c_0\left( 16\pi G \phi\right)^{1\over 2}\,\,,\,\, {\widetilde{\chi_{i}}}\equiv{\widetilde{N}}_{i}-c_i\left( 16\pi G \phi\right)\,\,,\,\, \nonumber\\
&&{\widetilde{\chi}}_4 \equiv {\widetilde{\pi}}_N\,\,,\,\, {\widetilde{\chi}}_{i+4} \equiv {\widetilde{\pi}}^i\,. 
\label{third1}
\end{eqnarray}
Imposing also here the gauge-fixing conditions \eqref{EFimpongo3} to be preserved, we have: 
\begin{equation}
\dot{\widetilde{\chi}}_0\approx\left\{\widetilde{N}(x)-c_0(16\pi G\phi(x))^{\frac{1}{2}},{{\widetilde{H}}_T^{'(-3/2)}}  \right\}\approx 0\,, 
    \label{condo3}
\end{equation}
which gets ${\widetilde{\lambda}}^{N}(x)\approx 0$,
and analogously from: 
\begin{equation}
\dot{\widetilde{\chi}}_i\approx\left\{{\widetilde{N}}_i(x)-c_i(16\pi G\phi(x)),{{\widetilde{H}}_T^{'(-3/2)}} \right\}\approx 0\,,
\label{condo2}
\end{equation}
it follows ${\widetilde{\lambda}}_{i}(x)\approx 0$.
Note that, also in this case, $\dot{\widetilde{\chi}}_4$ and $\dot{\widetilde{\chi}}_{i+4}$ 
are automatically preserved.

As usual, the second class Dirac's constraint matrices $C_{\alpha\beta}$ is: 
\begin{equation}
{\widetilde{C}}^{-1}_{\alpha\beta}\equiv 
 \left(
     \begin{array}{c c c c}
      0 & 0 & -1 & 0 \\
    0 & 0 &  0 & -\mathbb{I} \\
    1 & 0 & 0  &  0   \\
    0 & \mathbb{I} & 0 & 0
      \end{array}
    \right)\,. 
    \label{sectilde1}
\end{equation}
The equations of motion, calculated with Dirac's brackets and imposing strongly the second class constraints
are:
\begin{eqnarray}
\dot{\widetilde{h}}_{ij}&\approx&\{{\widetilde{h}}_{ij},{{\widetilde{H}}_T^{'(-3/2)}} \}_{DB}\nonumber\\
 &\approx& 16 \pi G\left(c_j {\widetilde{D}}_i \phi
+c_i  {\widetilde{D}}_j \phi\right)\nonumber\\
&&+ \frac{(32\pi G) c_0\left(16\pi G \phi\right)^{1\over 2}}{ \sqrt{\widetilde{h}}} \left({\widetilde{\pi}}_{ij}-\frac{{\widetilde{\pi}}_h}{2} \widetilde{h}_{ij}\right),
\label{metricoc3}
\end{eqnarray}
\begin{eqnarray}
\dot{\widetilde{\pi}}^{ij}&\approx&\{{\widetilde{\pi}}^{ij},{{\widetilde{H}}_T^{'(-3/2)}} \}_{DB}\nonumber\\
 &\approx&
 -\frac{c_0\left(16\pi G \phi\right)^{1\over 2} \sqrt{\widetilde{h}}}{16\pi G}\left ({}^{(3)}{\widetilde{R}}^{ij}-\frac{1}{2}{}^{(3)}{\widetilde{R}}\,{\widetilde{h}}^{ij}\right) \nonumber \\ 
 &&+\frac{(16\pi G) c_0\left(16\pi G \phi\right)^{1\over 2}}{2\sqrt{\widetilde{h}}}\widetilde{h}^{ij}\left( {\widetilde \pi}^{pk}{\widetilde \pi}_{pk}-\frac{{{\widetilde \pi}_h}^2}{2}\right) \nonumber\\
&&-2\frac{(16\pi G) c_0\left(16\pi G \phi\right)^{1\over 2}}{\sqrt{\widetilde{h}}}\left(\widetilde{\pi}^{ik}\widetilde{\pi}_k^{~j}-\frac{1}{2}\widetilde{\pi}_h \widetilde{\pi}^{ij}\right) \nonumber\\
&&-\widetilde{h}^{ij}\frac{\sqrt{{\widetilde h}} c_0\left(16\pi G \phi\right)^{1\over 2}}{2}V(\phi)\nonumber \\
&&+\frac{c_0\sqrt{\widetilde{h}}}{2\left(16\pi G\phi\right)^{1\over 2}}
\left[-\frac{\widetilde{D}^i \phi \widetilde{D}^j\phi}{2\phi}+\widetilde{D}^i\widetilde{D}^j \phi\right.\nonumber\\
&&\left. - \widetilde{h}^{ij}\left(-\frac{\widetilde{D}^k \phi \widetilde{D}_k\phi}{2\phi}+\widetilde{D}^k\widetilde{D}_k \phi\right)\right]\nonumber\\
&&+(16\pi G)\sqrt{\widetilde{h}} c^k \widetilde{D}_k \left(\frac{1}{\sqrt{\widetilde{h}}} \phi \widetilde{\pi}^{ij}\right)\nonumber\\ 
&&-(16\pi G)\widetilde{\pi}^{ki} c^j \widetilde{D}_k \phi
-(16\pi G)\widetilde{\pi}^{kj}c^i\widetilde{D}_k \phi \,, 
\label{piDBEF} 
\end{eqnarray}
\begin{eqnarray}
\dot{\phi}&\approx&\left\{\phi,\widetilde{H}_T^{\prime(-3/2)}\right\}_{DB}\approx -\widetilde{\lambda}_\phi \phi\,,
\end{eqnarray}
\begin{eqnarray}
\dot{\widetilde{\pi}}_\phi&\approx&\left\{\widetilde{\pi}_\phi,\widetilde{H}_T^{\prime(-3/2)}\right\}_{DB}\approx \widetilde{\lambda}_\phi {\widetilde{\pi}}_{\phi}.
    \label{piphidotEFeq1}
\end{eqnarray}

We stressed several times that, once we impose strongly the second class constraints, we can solve explicitly them and reduce the 
degrees of freedom of the phase space. 
Then, the transformation from the Jordan to the Einstein frame, on this reduced phase space, is Hamiltonian canonical transformation also in the case $\omega=-\frac{3}{2}$. As a corollary, we can map the equations of motion in the JF into the equations of motion in the EF (and vice-versa). 

%
% BibTeX users please use
\bibliographystyle{ieeetr}
%\bibliography{}
\bibliography{bransdickepartcase}
%%\begin{thebibliography}{}
%%
%% and use \bibitem to create references.
%Author, \textit{Book title} (Publisher, place year) page numbers
%\end{thebibliography}

\end{document}